%% file: 2023_08_31_Journal.tex
\documentclass[10pt,journal,twocolumn,romanappendices]{IEEEtran}
\IEEEoverridecommandlockouts

\usepackage{amsmath,amssymb,amsthm,bbm,cite,color,enumitem,graphicx,microtype,setspace,url}
\usepackage{graphicx}
\usepackage{tikz}
\usepackage[USenglish]{babel}
\usepackage[utf8]{inputenc}
\usepackage[T1]{fontenc}
\usepackage[algo2e]{algorithm2e}
\usepackage{algorithm}
\usepackage{comment}
\usepackage{epstopdf}
\usepackage{hyperref}
\usepackage{scrextend}
\SetKwComment{Comment}{/* }{ */}
\input{./Definitions.tex}

\newcommand{\tar}{\textnormal{\tiny{Tar}}}

\newcommand{\mse}{\mathrm{MSE}}
\newcommand{\sinr}{\mathrm{\gamma}}
\newcommand{\ue}{\textnormal{\tiny{UE}}}

\newcommand{\bmrc}{\textnormal{\tiny{BMRC}}}
\newcommand{\bmmse}{\textnormal{\tiny{BMMSE}}}
\newcommand{\rx}{\textnormal{\tiny{RX}}}

\newcommand{\low}{\textnormal{\tiny{low}}}
\newcommand{\high}{\textnormal{\tiny{high}}}
\newcommand{\pwr}{\textnormal{\tiny{$\rho$}}}
\newcommand{\npwr}{\textnormal{\tiny{$\sigma$}}}

\makeatletter
\renewcommand{\@IEEEsectpunct}{~}
\makeatother

\usepackage[labelsep=period,font=small]{caption}
\usepackage{tikz}
\usetikzlibrary{arrows, arrows.meta, backgrounds, calc, intersections, decorations.markings, decorations.pathreplacing, positioning, shapes}

\usepackage{pgfplots}
\usepgfplotslibrary{groupplots}
\pgfplotsset{compat=1.17}

\title{Uplink Transmit Power Optimization for Distributed Massive MIMO Systems with 1-Bit ADCs}
\author{Bikshapathi Gouda, Italo Atzeni, and Antti Tölli
\thanks{The authors are with the Centre for Wireless Communications, University of Oulu, Finland (e-mail: \{bikshapathi.gouda, italo.atzeni, antti.tolli\}@oulu.fi). This work is supported by the Research Council of Finland (336449 Profi6, 346208 6G~Flagship, 343586 CAMAIDE and 348396 HIGH-6G) and by the European Commission (101095759 Hexa-X-II). Part of this work was presented at IEEE GLOBECOM 2023~\cite{Gou23}.}\vspace{-5mm}}
\begin{document}

\maketitle
\begin{abstract}
This paper addresses the problem of uplink transmit power optimization in distributed massive multiple-input multiple-output systems, where remote radio heads (RRHs) are equipped with 1-bit analog-to-digital converters (ADCs). First, in a scenario where a single RRH serves a single user equipment (UE), the signal-to-noise-and-distortion ratio (SNDR) is shown to be a non-monotonic and unimodal function of the UE transmit power due to the quantization distortion (QD). Upon the introduction of multiple RRHs, adding properly tuned dithering at each RRH is shown to render the SNDR at the output of the joint receiver unimodal. In a scenario with multiple RRHs and UEs, considering the non-monotonic nature of the signal-to-interference-plus-noise-and-distortion ratio (SINDR), both the UE transmit powers and the RRH dithering levels are jointly optimized subject to the min-power and max-min-SINDR criteria, while employing Bussgang-based maximum ratio combining (BMRC) and minimum mean squared error (BMMSE) receivers. To this end, gradient and block coordinate descent methods are introduced to tune the UE transmit powers, whereas a line search coupled with gradient updates is used to adjust the RRH dithering levels. Numerical results demonstrate that jointly optimizing the UE transmit power and the RRH dithering levels can significantly enhance the system performance, thus facilitating joint reception from multiple RRHs across a range of scenarios. Comparing the BMMSE and BMRC receivers, the former offers a better interference and QD alleviation while the latter has a lower computational complexity.


\end{abstract}

\section{Introduction}

Fully digital massive multiple-input multiple-output (MIMO) is widely recognized for its ability to realize flexible beamforming and large-scale spatial multiplexing~\cite{Mar10}. As the next-generation wireless systems are moving towards higher carrier frequencies, both the bandwidth and the number of antennas at the base station (BS) are bound to increase significantly~\cite{Raj20}. However, fully digital massive MIMO arrays with high-resolution analog-to-digital converters (ADCs) are exceedingly complex and power intensive~\cite{Yan19}. To mitigate this issue, low-resolution and 1-bit ADCs can be employed, which significantly reduces the power consumption without excessively compromising the performance~\cite{Mo15, Jac17, Mol17, Li17, Atz22}. Furthermore, adopting low-resolution and 1-bit ADCs in a distributed (also known as cell-free) massive MIMO setting allows to reduce the backhaul signaling required for the data and channel state information (CSI) exchange among the BSs~\cite{Ngo17, Atz21, Gou24}.

For a single-BS massive MIMO system with low-resolution ADCs, the spectral and energy efficiency were characterized with respect to the transmit power of the user equipments (UEs) in~\cite{Pir18, Liu21}. On the other hand, studies such as~\cite{Jac17, Mol17, Li17} investigated the uplink channel estimation and the achievable signal-to-interference-plus-noise-and-distortion ratio (SINDR) with 1-bit ADCs. All the aforementioned works provided an asymptotic analysis of the SINDR assuming a Gaussian approximation of the quantization distortion (QD) arising from the 1-bit ADCs. This approximation is accurate in scenarios characterized by either a low UE transmit power or a large number of UEs~\cite{Mo15}. The evaluation of the error vector magnitude against the UE transmit power with few UEs indicated inferior performance with high UE transmit powers, which can be attributed to the non-Gaussian QD~\cite{Jac19}. However, appropriately tuned dithering (i.e., intentionally applied noise) at the BS can enhance the performance in such a scenario~\cite{Jac19}. A similar behavior was observed while examining the mean square error (MSE) of the channel estimation and the symbol error rate (SER) in~\cite{Atz22, Rad23}. These works demonstrated that the MSE of the channel estimation and the SER obtained with 16-QAM data symbols can be enhanced by appropriately tuning the UE transmit powers. To this end, an adaptive transmit power optimization mechanism tailored for a single-UE system with 1-bit ADCs was developed in~\cite{Rav23} to optimize the UE transmit power in order to minimize the SER. While many previous studies have considered conventional maximum ratio combining (MRC) or zero-forcing (ZF) receivers, it was shown in~\cite{Ngu21} that Bussgang-based MRC, ZF, and minimum mean square error (MMSE) receivers perform better, as they incorporate the Bussgang gain into the design.

The optimization of the uplink and downlink transmit powers enhances the network energy efficiency and ensures that each UE receives the required quality of service (QoS), while also mitigating the impact of near-far scenarios. In a coordinated multi-cell setting, the joint optimization of the UE transmit powers and the receivers at the BS under QoS constraints was addressed in~\cite{Ras98}. Assuming MMSE receivers,~\cite{Guo14} focused on optimizing the UE transmit power for both the pilot and data transmissions under per-UE QoS constraints. A similar study was provided in~\cite{Van18}, which also addressed the power allocation for pilot design to minimize the pilot contamination.  
Considering the complexities and scalability of global power optimization in multi-cell systems, a simplified approach was proposed in~\cite{Zha15} assuming constant interference from the other cells. Furthermore, a simplified transmit power optimization method via geometric mean considering max-min fairness in each cell was proposed in~\cite{Gha20}. In a multi-cell context with low-resolution ADCs, the optimization of the uplink and downlink transmit powers and beamformers is studied in~\cite{Cho21, Cho24} while assuming the Gaussian approximation for the QD.

In the context of distributed massive MIMO systems with joint transmission or reception, the optimization of the UE transmit powers and the receivers at the BSs to ensure fairness among the UEs was studied in~\cite{Bas18}. Furthermore, the uplink spectral efficiency was analyzed considering the UE transmit powers with both the ZF and MRC receivers in~\cite{Mai19}. In~\cite{Con22}, a scalable and less complex uplink transmit power optimization scheme was proposed under the max-min fairness criterion considering the MRC receiver. The energy efficiency of cell-free systems with low-resolution ADCs was analyzed in conjunction with an uplink power control scheme for max-min fairness in~\cite{Hu19, Zha21}. Additionally, the spectral efficiency of cell-free systems with 1-bit ADCs, considering quantization and channel estimation errors, was evaluated in~\cite{Zha19}. It is important to note that the works in~\cite{Hu19, Zha19, Zha21} also employ the Gaussian approximation for QD, which becomes increasingly accurate with higher ADC resolution. However, this approximation does not fully capture the true performance characteristics of 1-bit ADCs due to the non-Gaussian QD, especially with a low-to-moderate number of UEs.

\smallskip


\textit{Contribution.} In this paper, we examine the problem of uplink transmit power optimization in a distributed massive MIMO system where the remote radio heads (RRHs) are equipped with 1-bit ADCs. First, we provide insights into the signal-to-noise-and-distortion ratio (SNDR) of a UE in single- and multi-RRH scenarios. Our analysis reveals that the SNDR is non-monotonic and highly dependent on the UE transmit powers and locations due to the non-Gaussian QD of the 1-bit ADCs. Specifically, the SNDR is unimodal in the single-RRH case and multimodal in the multi-RRH case. Nonetheless, the SNDR can be made unimodal even in the multi-RRH scenario by tuning the dithering levels at the RRHs.\footnote{Radio frequency (RF) circuitry akin to automatic gain control can be used to tune the dithering levels at the RF stage before the 1-bit ADCs.} The joint optimization of the UE transmit powers and RRH dithering levels is tackled considering the min-power and max-min-SINDR objectives in a multi-UE, multi-RRH scenario employing the Bussgang-based MRC (BMRC) and MMSE (BMMSE) receivers from~\cite{Ngu21}. For a given set of RRH dithering levels, the UE transmit powers are optimized for both objectives using either the gradient or the block coordinate descent (BCD) method. The BCD method is augmented with a tailored termination criterion to prevent the UE transmit powers from entering the QD-dominated region of the SINDR. 
Additionally, we propose a simple approach for updating the RRH dithering levels based on a line search followed by a fine-tuning with gradient updates. Numerical results indicate that optimizing the UE transmit powers alongside the RRH dithering levels can greatly enhance the system performance by enabling joint reception from multiple RRHs across various scenarios. We further show that, if achieving the desired target SINDR requires reception over a large virtual antenna array resulting from aggregating multiple RRHs in combination with RRH-specific dithering, then the UE transmit power is heavily dominated by the distance to the farthest serving RRH.

The contributions of this paper are summarized as follows:
\begin{itemize}
    \item The UE transmit powers and RRH dithering levels are optimized for the considered distributed MIMO system with 1-bit ADCs taking into account the non-monotonic behavior of the SINDR due to the QD.
    \item Gradient and BCD methods are proposed to optimize the UE transmit powers for the min-power and max-min-SINDR objectives, considering both the BMRC and BMMSE receivers.
    \item A simple approach for updating the RRH dithering levels based on a line search is provided. Additionally, the RRH dithering levels are fine-tuned using the gradient method for the min-power and max-min-SINDR objectives. 
    \item Simulation results demonstrate that optimizing the UE transmit powers alongside the RRH dithering levels improves the system performance by facilitating joint reception from multiple RRHs in several scenarios.
\end{itemize}

Part of this work is included in our conference paper~\cite{Gou23}, which presented the UE transmit power optimization with fixed RRH dithering levels.

\smallskip


\textit{Outline.} The rest of the paper is structured as follows. Section~\ref{sec:SM} introduces the distributed massive MIMO system model with RRHs equipped with 1-bit ADCs. Section~\ref{sec:SINDRChar} characterizes the SINDR of a UE in both single and multi-RRH scenarios. Based on this, Sections~\ref{sec:min_pwr} and~\ref{sec:max_min_SINDR} delve into the optimization of the UE transmit powers and RRH dithering levels for the min-power and max-min-SINDR designs, respectively. Finally, Section~\ref{sec:NUM} presents the numerical results and Section~\ref{sec:CONC} provides some concluding remarks.

\smallskip


\textit{Notation.} Lowercase and uppercase boldface letters denote vectors and matrices, respectively. $(\cdot)^{\tran}$ and $(\cdot)^{\herm}$ represent the transpose and Hermitian transpose operators, respectively. $\|\cdot\|$ denotes the Euclidean norm for vectors. $\Re[\cdot]$, $\Im[\cdot]$, and $\Exp[\cdot]$ denote the real part, imaginary part, and expectation operators, respectively. $\mathbf{I}_{L}$ stands for the $L$-dimensional identity matrix. $\Diag(\cdot)$ produces a diagonal matrix with the elements of the vector argument or the diagonal elements of the square matrix argument on its diagonal. The Kronecker product is denoted by $\otimes$. $\mathrm{sgn}(\cdot)$ is the sign function. $\succ$ denotes the element-wise $>$ operator of a vector. $[a_{1}, \ldots, a_{L}]$ denotes the horizontal concatenation, whereas $\{a_{1}, \ldots, a_{L}\}$ and $\{ a_{\ell} \}_{\ell \in \setL}$ represent sets; the latter notation is occasionally relaxed as $\{ a_{\ell} \}$ for brevity. $\setC \setN(0, \sigma^{2})$ represents the complex normal distribution with zero mean and variance $\sigma^{2}$. $\frac{\partial}{\partial x}(\cdot)$ denotes the partial derivative with respect to $x$. $\mathcal{L}_{(\textrm{P})}(\cdot)$ represents the Lagrangian of the optimization problem~$(\textrm{P})$.

\begin{figure*}
\setcounter{equation}{13}
\begin{align}\label{eq:bar_sinr_k}
    \gamma_k \approx \hat{\sinr}_k \! \triangleq \! \frac{\rho_k |\w_k^{\herm} \hat \A \hat \h_k |^2}{\sum_{ k \in \setK} \rho_{ k} \|\w_k^{\herm} \hat \A \C_{\tilde \h_k}^{\frac{1}{2}}\|^2 \! + \sum_{\bar k \neq k} \rho_{\bar k} |\w_k^{\herm} \hat \A \hat \h_{\bar k} |^2 + \! \| \w_k^{\herm} \hat \A \C_{\z}^{\frac{1}{2}} \|^2 \! + \! \w_k^{\herm} \C_{\hat \q} \w_k} =  \frac{\rho_k |\w_k^{\herm} \hat \A \hat \h_k |^2}{\w_k^{\herm} \C_{\hat \r} \w_k - \rho_k |\w_k^{\herm} \hat \A \hat \h_k |^2}
\end{align}
\setcounter{equation}{0}
\hrule
\vspace{-3mm}
\end{figure*}

\section{System Model}\label{sec:SM}

We consider an uplink distributed massive MIMO system where a set of RRHs $\setB \triangleq \{1, \ldots, B\}$, each with $M$ antennas, serves a set of single-antenna UEs $\setK \triangleq \{1, \ldots, K\}$. Each RRH antenna is connected to two 1-bit ADCs, which quantize the in-phase and quadrature components of the received signal. Let $\h_{b,k} \in \Compl^{M \times 1}$ denote the uplink channel between UE~$k \in \setK$ and RRH~$b \in \setB$, where $\h_{k} \triangleq [\h_{1,k}^{\tran}, \ldots, \h_{B,k}^{\tran}]^{\tran} \in \Compl^{B M \times 1}$ is the aggregated uplink channel of UE~$k$ across all the RRHs. We use $d_k \sim \setC \setN (0, 1)$ to denote the data symbol transmitted by UE~$k$, which is independent from the data symbols of the other UEs. The received signal at RRH~$b$ prior to the 1-bit ADCs is given by
\begin{align}
\y_b \triangleq \sum_{k \in \setK} \sqrt{\rho_k} \h_{b,k} d_k + \z_b \in \Compl^{M \times 1},
\end{align}
where $\rho_k$ is the transmit power of UE~$k$ and $\z_b \sim \setC \setN (0, \sigma_{b}^{2} \I_{M})$ is the additive white Gaussian noise (AWGN) vector at RRH~$b$. Then, the received signal $\y_b$ is quantized as
\begin{align}\label{eq:r_b}
\r_b \triangleq Q(\y_b) \in \Compl^{M \times 1},
\end{align}
where $Q(\mathbf{a}) \triangleq \frac{1}{\sqrt{2}}\big(\sgn \big(\Re[\mathbf{a}]\big) + j \, \sgn \big(\Im[\mathbf{a}]\big)\big)$. 

Let us introduce $\y \triangleq [\y_1^{\tran}, \ldots, \y_B^{\tran}]^{\tran} \in \Compl^{BM \times 1}$ and $\z \triangleq [\z_1^{\tran}, \ldots, \z_B^{\tran}]^{\tran} \in \Compl^{BM \times 1}$ as the aggregated received signal and AWGN vector, respectively, across all the RRHs. Assuming perfect CSI, we can describe the 1-bit quantized output across all the RRHs using the Bussgang decomposition as~\cite{Bus52}
\begin{align} \label{eq:r_vec}
\r \triangleq Q(\y) = \A \y + \q \in \Compl^{BM \times 1},
\end{align}
where $\q$ is the zero-mean, non-Gaussian QD vector that is uncorrelated with $\y$ and
\begin{align}
    \A \triangleq \sqrt{\frac{2}{\pi}}\Diag(\C_{\y})^{-\frac{1}{2}} \in \Compl^{BM \times BM}
\end{align}
is the Bussgang gain, with $\C_{\y} \triangleq \Exp[\y\y^{\herm}] = \sum_{k \in \setK} {\rho_k} \h_{k} \h_k^{\herm} + \C_\z \in \Compl^{BM \times BM}$ and $\C_\z \triangleq \Diag \big([\sigma^2_1, \ldots, \sigma^2_B]\big) \otimes \I_{M}$.

The soft-detected symbol of UE~$k$ is obtained by combing $\r$ in \eqref{eq:r_vec} with the network-wide receiver $\w_k \in \Compl^{BM \times 1}$, i.e.,
\begin{align}
    \hat d_k & = \w_k^{\herm} \r \\
             & = \sqrt{\rho_k} \w_{k}^{\herm} \A \h_{k} d_k \! + \! \sum_{\bar k \neq k} \! \sqrt{\rho_{\bar k}} \w_{k}^{\herm} \A \h_{\bar k} d_{\bar k} \! + \! \w_{k}^{\herm} \A \z \! + \! \w_{k}^{\herm} \q.
\end{align}
Then, the corresponding SINDR of UE~$k$ is given by
\begin{align}\label{eq:sinr_k}
    \sinr_k \! & \triangleq \! \frac{\rho_k |\w_k^{\herm} \A \h_k |^2}{\sum_{\bar k \neq k} \rho_{\bar k} |\w_k^{\herm} \A \h_{\bar k} |^2 \! + \! \| \w_k^{\herm} \A \C_{\z}^{\frac{1}{2}} \|^2 \! + \! \w_k^{\herm} \C_{\q} \w_k} \\
    & = \! \frac{\rho_k |\w_k^{\herm} \A \h_k |^2}{\w_k^{\herm} \C_{\r} \w_k - \rho_k |\w_k^{\herm} \A \h_k |^2 },
\end{align}
where  $\C_{\q} \triangleq \Exp [\q \q^{\herm}] = \C_\r - \A \C_{\y} \A^{\herm} \in \Compl^{BM \times BM}$ and~\cite{Li17}
\begin{align}\label{eq:cr}
    \C_\r  & \triangleq \frac{2}{\pi} \big( \arcsin \big( \Diag(\C_{\y})^{-\frac{1}{2}} \Re[\C_\y] \Diag(\C_{\y})^{-\frac{1}{2}} \big) \nonumber \\ 
    & \phantom{=} \ + j \, \arcsin \big( \Diag(\C_{\y})^{-\frac{1}{2}}\Im[\C_\y] \Diag(\C_{\y})^{-\frac{1}{2}} \big) \big).
\end{align}

In the following section, we present an approximation of the SINDR using the estimated channels at the RRHs. Subsequently, we explore the SNDR of a single UE with RRHs equipped with 1-bit ADCs. Note that, in the absence of interference from other UEs, the terms SINDR and SNDR can be used interchangeably.

\section{Characterization of the SINDR} \label{sec:SINDRChar}

Let us define $\hat{\h}_k \triangleq \h_k - \tilde{\h}_k$ as the channel estimate of UE~$k$ across all the RRHs, where $\tilde{\h}_k$ denotes the channel estimation error. We assume that $\tilde{\h}_k$ is independent of $\hat{\h}_k$, with $\Exp[\tilde{\h}_k] = \0$ and $\Exp[\tilde{\h}_k \tilde{\h}^{\herm}_k] \triangleq \C_{\tilde{\h}_k}$~\cite{Wan18, Yan24}. With these definitions, the estimate of $\C_{\y} $ can be written as $\hat \C_{\y} \triangleq \sum_{k \in \setK} {\rho_k} (\hat \h_{k} \hat \h_k^{\herm} + \C_{\tilde \h_k})  + \C_\z$.  Consequently, the estimate of $\A$ is given by 
\begin{align}
   \hat  \A \triangleq \sqrt{\frac{2}{\pi}}\Diag( \hat \C_{\y})^{-\frac{1}{2}},
\end{align}
and the quantized received signal in \eqref{eq:r_vec} is approximated as
\begin{align} \label{eq:r_bar_vec}
\r \approx \hat{\r} \triangleq \hat{\A} {\y} + \hat{\q},
\end{align}
where $\hat{\q}$ represents the approximated QD.

After combining $\hat \r$ in \eqref{eq:r_bar_vec} with the receiver $\w_k$, the soft-detected symbol is approximated as
\begin{align}\label{eq:d_k_app}
   \hat d_k &  \approx \w_k^{\herm} \hat \r \\
             & = \sqrt{\rho_k} \w_{k}^{\herm} \hat \A \hat \h_{k} d_k 
             + \sum_{k \in \setK} \sqrt{\rho_k} \w_{k}^{\herm} \hat \A \tilde \h_{k} d_k  \nonumber \\
             & \phantom{=} \ + \sum_{\bar k \neq k}  \sqrt{\rho_{\bar k}} \w_{k}^{\herm} \hat \A \hat \h_{\bar k} d_{\bar k} 
             + \w_{k}^{\herm} \hat \A \z \! + \! \w_{k}^{\herm} \hat \q.
\end{align}
Considering the channel estimation error in \eqref{eq:d_k_app}, the SINDR of UE~$k$ is approximated as shown in \eqref{eq:bar_sinr_k} at the top of the page, \setcounter{equation}{14} with $\C_{\hat \q} \triangleq \Exp [\hat \q \hat \q^{\herm}] = \C_{\hat \r} - \hat \A \hat \C_{\y} \hat \A^{\herm}$ and where $\C_{\hat \r}$ is obtained by replacing $\C_{\y}$ with $ \hat \C_{\y}$ in \eqref{eq:cr}. As $\hat \h_k$ approaches $\h_k$, the SINDR $\hat \gamma_k$ in \eqref{eq:bar_sinr_k} converges to $\gamma_k$ in \eqref{eq:sinr_k}.

From the estimated channels, the BMRC receiver of UE~$k$ is defined as 
\begin{align}\label{eq:bmrc_wk}
    \w_k^{\bmrc} \triangleq \hat \A \hat \h_k,
\end{align}
while the BMMSE receiver of UE~$k$ is obtained by minimizing its MSE, defined as $\mse_k \triangleq \Exp \big[ |(\w_k^{\bmmse})^{\herm} \hat \r - d_k|^2 \big]$, which yields
\begin{align}\label{eq:bmmse_wk}
\w_k^{\bmmse} = \sqrt{\rho_k} \C_{\hat \r}^{-1} \hat \A \hat \h_k.
\end{align}
Furthermore, the SINDR in \eqref{eq:bar_sinr_k} is rewritten for the BMRC and BMMSE receivers as
\begin{align}\label{eq:bmrc_sinr_k}
    \hat \sinr^{\bmrc}_k & \triangleq  \frac{\rho_k | \hat \h_k ^{\herm} \hat \A  \hat \A \hat \h_k |^2}{\hat \h_k ^{\herm} \hat \A  \C_{\hat \r}  \hat \A \hat \h_k- \rho_k | \hat \h_k ^{\herm} \hat \A \hat \A \hat \h_k |^2}, \\\label{eq:bmmse_sinr_k}
    \hat \sinr^{\bmmse}_k & \triangleq  \frac{\rho_k | \hat \h_k ^{\herm} \hat \A \C^{-1}_{\hat \r} \hat \A \hat \h_k |^2}{\hat \h_k ^{\herm} \hat \A  \C^{-1}_{\hat \r}  \hat \A \hat \h_k- \rho_k | \hat \h_k ^{\herm} \hat \A \C^{-1}_{\hat \r} \hat \A \hat \h_k |^2},
\end{align}
respectively.

\subsection{SNDR in the Single-UE Case}

In this section, we explore the SNDR behavior of a single UE in single- and two-RRHs scenarios. For the SNDR analysis, we assume $M=128$ antennas and the BMMSE receiver with perfect CSI at each RRH.\footnote{A similar SNDR behavior can be observed with the BMRC receiver. It is important to note that, even for a single UE, the BMMSE receiver is not a scaled version of the BMRC receiver due to the non-linear terms in $\C_{\r}$ in \eqref{eq:cr}, which restricts the use of the Woodbury matrix identity as in the unquantized case.} Furthermore, we model the channel between UE~$k$ and RRH~$b$, denoted by $\h_{b,k} \in \Compl^{M \times 1}$, using uncorrelated Rayleigh fading, i.e., $\h_{b,k} \sim \setC \setN (0,\delta_{b,k} \I_{M})$. The large-scale fading coefficient $\delta_{b,k}$ is determined by the distance $d_{b,k}$ between UE~$k$ and RRH~$b$, given by $\delta_{b,k} \triangleq -61 - 30\log_{10}(d_{b,k})$ [dB], which assumes a $28$~GHz carrier frequency and a pathloss exponent of $3$.

\smallskip

\subsubsection{Single-RRH case: SNDR versus UE transmit power.}


\begin{figure}[t]
    \centering
    \includegraphics[scale=0.3]{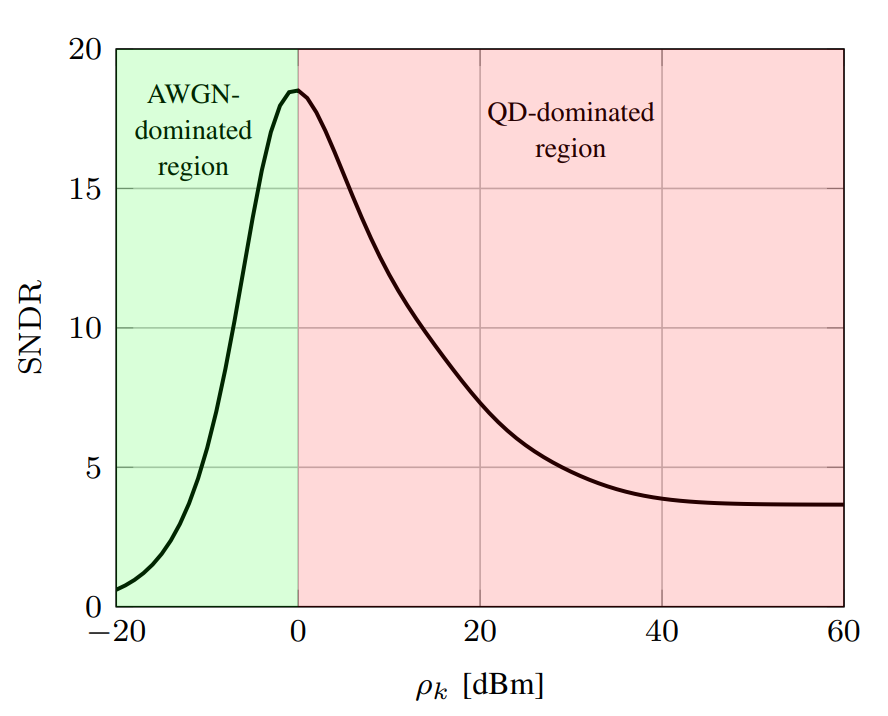}
\caption{Single-RRH case: SNDR versus UE transmit power.}
\label{fig:sinBSSINDR}
\end{figure}

 Let us begin with a scenario featuring a single RRH equipped with 1-bit ADCs, where the UE is positioned at a fixed distance of $30$~m from the RRH. Fig.~\ref{fig:sinBSSINDR} illustrates the SNDR as a function of the UE transmit power. The SNDR exhibits a non-monotonic yet unimodal behavior, with an optimal value achieved at a specific UE transmit power (around 0~dBm). For lower UE transmit powers, the impact of the AWGN dominates, overpowering the intended signal. Conversely, for higher UE transmit powers, the soft-estimated symbols suffer from amplitude loss due to severe clipping, resulting in the QD prevailing over the intended signal~\cite{Atz22}. Therefore, achieving the optimal SNDR requires an appropriate amount of UE transmit power that minimizes the combined effect of both AWGN and QD. 
 In the general case, the non-monotonic behavior of the SINDR for UE~$k$ due to QD is described in the following definition.

\begin{definition}\label{def:nonmonQD}
The uplink SINDR of UE~$k$, denoted by $\gamma_k$, exhibits non-monotonic behavior with respect to the UE transmit power due to the QD if there exists a UE transmit power $\rho_k > \rho'_k$ for which $\gamma_k(\rho_k) < \gamma_k(\rho'_k)$ for fixed interference and AWGN at the RRHs.
\end{definition}

\subsubsection{Two-RRH case: SNDR versus distance to the reference RRH.}\label{subsec:2BSsDis}

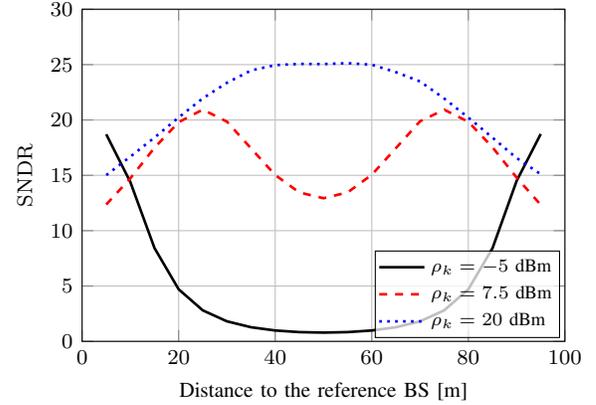
\begin{figure}[t]
   \centering
            \input{FigJor/fig2.tex}
            \vspace{-1mm}
            \caption{Two-RRH case: SNDR versus distance to the reference RRH.} \label{fig:2BSSINDR}
            \vspace{-3mm}
        \end{figure}
     
The SNDR of a UE in a two-RRH scenario can be characterized as a function of the distance between the UE and one of the two RRHs, which is referred to as the \textit{reference RRH} in Fig.~\ref{fig:2BSSINDR}. In this scenario, the two RRHs are positioned $100$~m apart. The UE maintains a fixed transmit power and moves along the line connecting the two RRHs. For a low UE transmit power of $-5$~dBm, the SNDR decreases as the UE moves away from the left RRH and increases as it gets closer to the other RRH as in the unquantized case, where the SNDR is primarily affected by the AWGN. For a high UE transmit power of $7.5$~dBm, the SNDR is limited by the QD of the closest RRH and the AWGN of the farthest RRH. However, as the UE moves away from the closest RRH, the detrimental impact of both the QD and AWGN is reduced, thereby improving the SNDR. At the cell edge, i.e., at $50$~m, the SNDR is dominated by the AWGN at both RRHs due to the limited UE transmit power, resulting in a lower SNDR. At a very high UE transmit power of $20$~dBm, the SNDR is limited by the QD if the UE is close to any of the RRHs, and the impact of the QD reduces as the UE moves towards the cell edge, resulting in a better SNDR. Note that the UE is not power-limited even at the cell edge in this scenario.

\smallskip

\subsubsection{Two-RRH case: SNDR versus UE transmit power.}
   \begin{figure}[t]
   \centering
            \input{FigJor/fig3.tex}
            \vspace{-1mm}
            \caption{Two-RRH case: SNDR versus UE transmit power.} \label{fig:2BSSINDRAWGNtun}
            \vspace{-3mm}
        \end{figure}
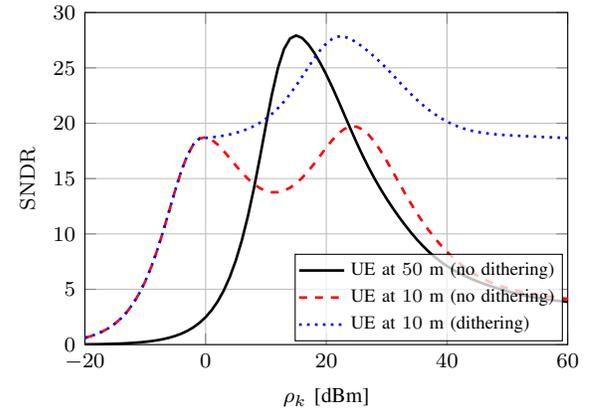
Let us analyze the impact of the UE transmit power on the SNDR for a single UE at a fixed distance from both RRHs. 
The impact of dithering (i.e., intentionally applied noise) at each RRH is also evaluated. If the UE is equidistant from both RRHs, specifically located $50$~m away from the reference RRH, the impact of the AWGN and QD on the SNDR is the same at both RRHs for a fixed UE transmit power. Similar to the single-RRH case, the optimal performance at both the RRHs is achieved at the same UE transmit power, leading to a unimodal SNDR behavior, as shown in Fig.~\ref{fig:2BSSINDRAWGNtun}. Moreover, the SNDR in the two-RRH scenario is higher than in the single-RRH case (see Fig.~\ref{fig:sinBSSINDR}) due to the combining gain from both RRHs. On the other hand, if the UE is closer to one of the RRHs (say $10$~m), the optimal performance at the closest RRH is achieved with a very low UE transmit power compared with the farthest RRH. This results in a bimodal SNDR behavior with two local maxima. 
Therefore, to enhance the overall SNDR, the dominance of the QD at the closest RRH can be reduced by adjusting either the dithering level or both the signal and dithering levels at each RRH~\cite{Sun10, Lei19}. 
For simplicity, we consider adjusting only the dithering level at each RRH throughout the paper. As a result, $\sigma_b$ becomes a controllable parameter referred to as the \textit{dithering level}, consisting of a superposition of both the AWGN and the additional Gaussian dithering signal introduced at each RRH. 
By adjusting the dithering level, a fixed QD level can be maintained (as mentioned in Remark~\ref{rm:qd}), resulting in an improved SNDR with unimodal behavior, as shown in Fig.~\ref{fig:2BSSINDRAWGNtun}. In scenarios involving over two RRHs and a UE with unlimited transmit power, the dithering levels can be applied to all the RRHs except the farthest one (constrained by the AWGN) to improve the SNDR. In practice, the optimal dithering level at each RRH depends on its pathloss ratio to the farthest RRH.

\begin{remark}\label{rm:qd}
By adjusting 
both the signal and dithering levels at each RRH, the QD can be kept constant regardless of variations in the UE transmit powers.
\end{remark}
Considering a single UE and neglecting the UE index, 
the quantized received signal at RRH~$b$ given in \eqref{eq:r_b} can be simplified as
\begin{align}\label{eq:r_b_sim}
\r_b = Q\bigg(\sqrt{\rho \delta_{b}}\bigg(\h'_{b} d + \frac{\sigma_b}{\sqrt{\rho \delta_b }} \z_b'\bigg) \bigg),
\end{align}
with $\h'_{b}, \z_b' \sim \setC \setN (0, \I_{M})$ and where $\rho$ is the UE transmit power.

It is clear from \eqref{eq:r_b_sim} that, as long as the ratio $\frac{{\sigma_b}}{\sqrt{\rho \delta_b}}$ remains constant, the output of the 1-bit ADC is the same. Therefore, in general, $\C_{\r}$ in \eqref{eq:cr} remains unchanged. Moreover, it can be shown that $\A\y$ in \eqref{eq:r_vec} does not change if ${\sigma_b}$ and $\sqrt{\rho \delta_b}$ vary with a fixed ratio. Therefore, by fixing $\frac{{\sigma_b}}{\sqrt{\rho \delta_b}}$ for different input signal powers, $\C_{\q}$  (i.e., the QD) is ensured to remain constant. We recall that the value of $\frac{{\sigma_b}}{\sqrt{\rho \delta_b}}$ can be maintained by adjusting either the RRH dithering level ${\sigma_b}$ or both the signal power $\rho$ and the RRH dithering level ${\sigma_b}$.

\section{Uplink Transmit Power Optimization Based on Min-Power}\label{sec:min_pwr}

In this section, we discuss the optimization of UE transmit powers to achieve a fixed target SINDR. Additionally, we can also optimize the dithering levels at the RRHs to attain a higher target SINDR, as discussed in Section~\ref{sec:SINDRChar}. 
For a given target SINDR $\gamma_k^{\tar}$ of UE~$k$, the min-power optimization problem can be formulated as
\begin{align}
\begin{array}{cl}
\label{eq:p1} \displaystyle \underset{\{\rho_{k}, \sigma_b\}}{\minimize} & \displaystyle \sum_{k \in \setK} \rho_{k} \\
\st & \hat \sinr_k \ge \gamma_k^{\tar}, \quad \forall k \in \setK \\
    & \sigma_{b} \ge \sigma_{\min}, \quad \forall b \in \setB,
\end{array}
\end{align}
where $\bar \sinr_k$ is given in \eqref{eq:bar_sinr_k} and $\sigma^2_{\min}$ is the minimum dithering level corresponding to the AWGN power. Problem \eqref{eq:p1} is not convex with respect to $\rho_{k}$ or $\sigma_b$.\footnote{Note that the optimization of $\sigma_{b}$ is not required for the single RRH case.} However, considering the SINDR properties discussed in Section~\ref{sec:SINDRChar}, the optimization of both the RRH dithering levels and the UE transmit powers in a multi-RRH scenario can be achieved, as discussed in the following. First, the optimization of the UE transmit powers is considered for a fixed set of dithering levels across all the RRHs, as discussed next.


\subsection{Optimization of the UE Transmit Powers}\label{sec:min_pwr_tp}


The SINDR constraint in \eqref{eq:p1} can be expressed using 
\eqref{eq:bmrc_sinr_k} or \eqref{eq:bmmse_sinr_k} as 
\begin{align}\label{eq:sindr_con_bmrc}
\hspace{-2mm}\xi^{\bmrc}_k & \!  \triangleq  \! (\gamma_k^{\tar} \! + \! 1) {\rho_k | \hat \h_k ^{\herm} \hat \A \hat \A \hat \h_k |^2} - \!\gamma_k^{\tar} {\hat \h_k ^{\herm} \hat \A \C_{\hat \r} \hat \A \hat \h_k} \! \ge 0, \\
\label{eq:sindr_con_bmmse}
\hspace{-2mm} \xi^{\bmmse}_k & \triangleq (\gamma_k^{\tar}+1)\rho_k \hat \h_k^{\herm} \hat \A \C_{\hat \r}^{-1} \hat \A \hat \h_k - {\gamma_k^{\tar}} \ge 0,
\end{align}
respectively. To further simplify the notation, we define $\mathrm{RX} \triangleq \{\mathrm{BMRC}, \mathrm{BMMSE}\}$ to represent the use of either the BMRC or the BMMSE receiver. For fixed RRH dithering levels $\{\sigma_b\}$, the SINDR of a UE with 1-bit quantization is a non-monotonic function of $\rho_k$ and inherently bounded, unlike the signal-to-interference-plus-noise ratio (SINR) in the unquantized case, even without interference from other UEs. Therefore, existing approaches such as~\cite{Ras98, Wie06} cannot be directly applied to optimize the UE transmit powers. Instead, we propose to use either the gradient or the BCD method for optimizing the UE transmit powers. 

\smallskip

\textit{\textbf{Gradient method.}} For a fixed set of RRH dithering levels $\{\sigma_b\}$, the Lagrangian of \eqref{eq:p1} for a given receiver is given by
\begin{align}
\label{eq:p2Lag_bmrc}
\mathcal{L}^{\rx}_{\eqref{eq:p1}}\big(\{\rho_k, \mu_k^{\rx}\}\big) & \triangleq \sum_{k \in \setK} \big(\rho_{k} - \mu_k^{\rx} \xi_k^{\rx} \big),
\end{align}
where $\mu_k^{\rx} \geq 0$ is the Lagrangian dual variable associated with the SINDR constraint in \eqref{eq:p1} for a given BMRC or BMMSE receiver. The transmit power of UE~$k$ is then updated iteratively in the direction of the negative gradient of the Lagrangian for a given receiver as
\begin{align}
\label{eq:gdUp_pwr}
\rho_k^{(i)} = \rho_k^{(i-1)} - \zeta_{\pwr} \frac{\partial }{\partial \rho_k}\mathcal{L}^{\rx}_{\eqref{eq:p1}} \big(\{\rho_k, \mu_k^{\rx}\} \big),
\end{align}
where $\zeta_{\pwr} \ge 0$ is the step size and $i$ is the iteration number.  For the initial iteration, $\{\rho_k\}$ are initialized with positive random numbers. The gradient of the Lagrangian with respect to $\rho_k$ is given in \eqref{eq:gradLag_bmrc} and \eqref{eq:gradLag_bmmse} at the top of the next page with the BMRC and BMMSE receivers, respectively, whereas the partial derivatives $\frac{\partial}{\partial \rho_{k}}\hat \A$ and $\frac{\partial}{\partial \rho_{ k}}  \C_{\hat \r} $ are provided in Appendix~\ref{app:rho_upd}. \setcounter{equation}{26} The Lagrangian dual variable $ \mu_k^{\rx}$ is updated at iteration $i$ as
\begin{align}\label{eq:mu_up_bmrc}
\mu_k^{\rx(i)} & = \mathrm{max}\big(0, \mu_k^{\rx(i-1)} - \nu \xi_k^{\rx} \big),
\end{align}
where $\nu \ge 0$ is the step size. For the initial iteration, $\{\mu_k^{\rx}\}$ are initialized with positive random numbers.

\begin{figure*}
\setcounter{equation}{24}
\begin{align}
\label{eq:gradLag_bmrc} \frac{\partial}{\partial \rho_k} \mathcal{L}^{\bmrc}_{\eqref{eq:p1}}\big(\{\rho_k, \mu_k^{\bmrc}\}\big) & = 1 - \mu_k^{\bmrc} (\gamma_k^{\tar}+1)  |\hat \h_k^{\herm} \hat \A \hat \A  \hat \h_k |^2 -  \sum_{\bar k \in \setK} \mu_{\bar k}^{\bmrc} \bigg( 4\rho_{\bar k} (\gamma_{\bar k}^{\tar}+1)  \hat \h_{\bar k}^{\herm} \hat \A \hat \A  \hat \h_{\bar k}   \Re \bigg[ \hat \h_{\bar k}^{\herm} \bigg(\frac{\partial}{\partial \rho_k}\hat \A \bigg) \hat \A  \hat \h_{\bar k} \bigg]  \nonumber \\ & \phantom{=} \ - 2 \gamma_k^{\tar}  \Re \bigg[\hat \h_{\bar  k}^{\herm} \bigg(\frac{\partial}{\partial \rho_{k}}\hat \A \bigg) \C_{\hat \r} \hat \A  \hat \h_{\bar  k} \bigg] -  \gamma_k^{\tar} \hat \h_{\bar k}^{\herm} \hat \A  \bigg( \frac{\partial}{\partial \rho_{ k}}  \C_{\hat \r}  \bigg) \hat \A  \hat \h_{ \bar k} \bigg)
\end{align}
\vspace{-2mm}
\hrule
\begin{align}
\label{eq:gradLag_bmmse} \frac{\partial}{\partial \rho_k} \mathcal{L}^{\bmmse}_{\eqref{eq:p1}}\big(\{\rho_k, \mu_k^{\bmmse}\}\big) & = 1 - (\gamma_k^{\tar}+1) \mu_k^{\bmmse}  \hat \h_k^{\herm} \hat \A \C_{\hat \r}^{-1} \hat \A  \hat \h_k  - \sum_{\bar k \in \setK} (\gamma_{\bar k}^{\tar}+1) \mu_{\bar k}^{\bmmse} \rho_{\bar k}  \bigg( \Re \bigg[\hat \h_{\bar  k}^{\herm} \bigg( 2 \frac{\partial}{\partial \rho_{k}}\hat \A \nonumber \\ & \phantom{=} \ -  \hat \A \C_{\hat \r}^{-1} \frac{\partial}{\partial \rho_{ k}}  \C_{\hat \r} \bigg) \C_{\hat \r}^{-1}\hat \A  \hat \h_{ \bar k}\bigg]\bigg)
\end{align}
\setcounter{equation}{27}
\vspace{-2mm}
\hrule
\vspace{-5mm}
\end{figure*}

Note that the gradient method in \eqref{eq:gdUp_pwr} terminates when feasible target SINDRs are met for all the UEs. However, to address the possibility of infeasible SINDR targets, a fixed number of iterations for the gradient method is imposed to guarantee a finite computation time. The convergence speed of this method depends on the slope of the non-monotonic SINDR function at different UE transmit powers. Moreover, in the case of multiple RRHs, this method may converge to a local optimum due to the presence of multiple peaks in the SINDR function, as discussed in Section~\ref{sec:SINDRChar}.

As an alternative solution to find an appropriate set of UE transmit powers to satisfy the SINDR constraints of \eqref{eq:p1}, a fixed-point power update method akin to~\cite{Ras98, Wie06} can be devised. This approach can be especially effective for low UE transmit powers, where the SINDR resembles the SINR of an unquantized system. Based on this, in the following, we propose a modified BCD method to optimize the UE transmit powers for the min-power objective.

\smallskip

\textit{\textbf{BCD method.}} In this method, the transmit power of a UE is updated with a scaling factor while the transmit powers of the other UEs are fixed in each iteration. For a given iteration $i$, a UE $k$ with the largest deviation from the SINDR target is selected to update its transmit power:
\begin{align}\label{eq:ue_idx}
k = \argmax_{\bar k \in \setK} {\gamma_{\bar k}^{\tar}}/{\hat \sinr_{\bar k}^{(i-1)}}.
\end{align}
Then, the transmit power of the chosen UE is scaled to satisfy the target SINDR constraint $\gamma_k^{\tar}$ 
as
\begin{align}
\label{eq:pwrScal} \rho_k^{(i)} = \min\big(\alpha, {\gamma_k^{\tar}}/{\hat \sinr_k^{(i-1)}}\big) \rho_k^{(i-1)},
\end{align}
where $\hat \sinr_k^{(i-1)}$ can be obtained using either the BMRC or BMMSE receiver as in \eqref{eq:bmrc_sinr_k} or \eqref{eq:bmmse_sinr_k}, respectively, given fixed transmit powers $\{\rho_k^{(i-1)}\}_{k \in \setK}$. The parameter $\alpha  > 1$ is used to limit the maximum power increase within a single iteration to prevent the UE transmit power from entering too far into the QD-dominated region of the SINDR function. It is important to note that the values of the UE transmit powers must be initialized such that the initial values of the SINDR are in the AWGN-dominated region. As the SINDR is non-monotonic, unlike in unquantized systems, we impose a termination criterion that stops the process if the SINDR of a UE starts decreasing when its transmit power is increased while the other UE transmit powers are fixed (i.e., $\gamma_k^{(i)}<\gamma_k^{(i-1)}$ if $\rho^{(i)}>\rho^{(i-1)}$), as described in Definition~\ref{def:nonmonQD}. Additionally, the process is terminated if the SINDR of all the UEs decreases when all the UE transmit powers are increased ($\gamma_k(\boldsymbol{\rho}) < \gamma_k(\boldsymbol{\rho'}),~\forall k \in \setK$ if $\boldsymbol{\rho} \succeq \boldsymbol{\rho'}$), as stated in Theorem~\ref{thm:qdscMinpwr}. The full procedure to optimize the UE transmit powers via the BCD method is summarized in Algorithm~\ref{alg:pwrScal}.

\begin{theorem}\label{thm:qdscMinpwr}
 The transmit power of a UE is in the QD-dominated region of the non-monotonic SINDR function if there exists a UE transmit power vector $\boldsymbol{\rho} \succeq \boldsymbol{\rho'}$ such that $\gamma_k(\boldsymbol{\rho}) < \gamma_k(\boldsymbol{\rho'}),~\forall k \in \setK$. Here, $\boldsymbol{\rho} = [\rho_1, \ldots, \rho_K]^{\tran}$ and $\boldsymbol{\rho'} = [\rho'_1, \ldots, \rho'_K]^{\tran}$.  
\end{theorem}

\begin{IEEEproof}
The theorem is proven by contradiction. Consider an unquantized system 
following the properties of standard interference functions~\cite{Yat95}. Let us begin with $\boldsymbol{\rho} \succeq \boldsymbol{\rho'}$, which implies that there exists a UE~$k$ for which $a \triangleq \frac{\rho_k}{\rho'_k} \ge \frac{\rho_{\bar{k}}}{\rho'_{\bar{k}}} \ge 1,~\forall \bar{k} \in \setK \setminus {k}$. This further implies $a \boldsymbol{\rho'} \succeq \boldsymbol{\rho}$.
Consequently, given a fixed receiver, the following condition holds for the SINR of UE~$k$:
\begin{align}
    \gamma_k(\boldsymbol{\rho'}) &=  \frac{G_k(\rho'_k)}{I_k(\boldsymbol{\rho'})} 
             \le  a\frac{G_k(\rho'_k)}{I_k(a\boldsymbol{\rho'})}
             \le \frac{G_k(\rho_k)}{I_k(\boldsymbol{\rho})} =  \gamma_k(\boldsymbol{\rho}),
\end{align}
where $G_k$ and $I_k$ are the signal and interference powers, respectively, of UE~$k$. The first and second inequalities follow from the scalability and monotonicity of the standard interference function~\cite{Yat95}. Therefore, 
if  $\boldsymbol{\rho} \succeq \boldsymbol{\rho'}$ 
, then $\gamma_k(\boldsymbol{\rho}) \ge \gamma_k(\boldsymbol{\rho}')$ for at least one UE. Consequently, our initial hypothesis of $\gamma_k(\boldsymbol{\rho}) < \gamma_k(\boldsymbol{\rho'}),~\forall k \in \mathcal{K}$ if $\boldsymbol{\rho} \succeq \boldsymbol{\rho'}$, necessitates the non-monotonic SINR behavior of a UE as described in Definition~\ref{def:nonmonQD}, attributable to the effects of the QD.
\end{IEEEproof}

\smallskip

\begin{figure}[t!]
\begin{minipage}[t]{0.9\linewidth}
\begin{algorithm}[H]
\footnotesize
\textbf{Data:} $\{\hat \h_k\}$, $\{\C_{\tilde \h_k}\}$, $\{\gamma_k^{\tar}\}$, $\{\sigma_b \}$. \\
\textbf{Initialization:} Iteration~$i=0$, $\{\rho_k^{(0)} \}$.
\begin{itemize}
\item Compute $\{\hat \sinr_k^{(0)}\}$ as in \eqref{eq:bmrc_sinr_k} or \eqref{eq:bmmse_sinr_k}.
\end{itemize}
\While{$ \underset{k \in \setK}{\min}(\hat \sinr_k^{(i)} - \gamma_k^{\tar}) < 0$}
{
    \begin{itemize}
    \item $i = i+1$.
    \item Select UE~$k$ according to \eqref{eq:ue_idx} to scale its transmit power.
    \item Update the UE transmit power as in \eqref{eq:pwrScal}.
    \item Compute $\{\hat \sinr_k^{(i)}\}$ as in \eqref{eq:bmrc_sinr_k} or \eqref{eq:bmmse_sinr_k}.
    \end{itemize}
        \If{ $\hat \sinr_k^{(i)} \! < \! \hat \sinr_k^{(i-1)} $  \textnormal{or} $ \underset{j \triangleq \{1, \ldots, i-1\}}{\max} \underset{\bar k \in \setK}{\min}(\hat \gamma_{\bar k}^{(i)} - \hat \gamma_{\bar k}^{(j)}) < 0$}  
        {
          \textbf{break}  \% To avoid the QD-dominated region of SINDR. 
        } 
  }
\caption{Optimization of the UE transmit powers via the BCD method} \label{alg:pwrScal}
\end{algorithm}
\end{minipage}
\vspace{-5mm}
\end{figure}

\smallskip

The proposed BCD method is particularly effective in the AWGN-dominated region, where the SINDR increases steadily and monotonically with the UE transmit powers. Additionally, the convergence speed is similar to the unquantized case and does not depend on the (fixed) step size, unlike the gradient method. However, when the optimization of RRH-specific dithering levels is added to the problem, a 
unique solution may not be always guaranteed due to the non-standard properties of interference and quantization distortion~\cite{Yat95}. 
In the following, we discuss the optimization of the dithering levels at the RRHs.


\subsection{Optimization of the RRH Dithering Levels}

As discussed in Remark~\ref{rm:qd}, the QD remains constant for a fixed $\frac{{\sigma_b}}{\sqrt{\rho \delta_b}}$ ratio. Therefore, when considering a fixed set of  $\{\sigma_b\}$ values, solving the UE transmit power optimization problem \eqref{eq:p1} may lead to a suboptimal or infeasible solution. In principle, we can approximately solve \eqref{eq:p1} by considering discretized sets of $\{\sigma_b\}$ values and choosing the best combination among them. However, the number of possible combinations of $\{\sigma_b\}$ values grows exponentially with the  increasing number of RRHs. 
In the following, we first present a heuristic approach to provide a coarse initial estimate of $\{\sigma_b\}$ using a simple line search. 
Then, a further refinement step is introduced, where the local optimal values close to the initial estimates are found using the gradient method.


\smallskip

\textit{\textbf{Coarse initialization of the RRH dithering levels.}}
The SINDR gain of a UE with joint reception from multiple RRHs increases as the ratios $\frac{{\sigma_b}}{\sqrt{\rho\delta_{b}}}$ become closer to each other. Consequently, the SINDR gain is maximum when all the RRHs have the same $\frac{{\sigma_b}}{\sqrt{\rho \delta_{b}}}$ (i.e., the QD is fixed at all the RRHs) as explained in Section~\ref{sec:SINDRChar}. In the multi-UE scenario, a fixed QD for all the UEs cannot be guaranteed at all the RRHs simultaneously. Moreover, in a distributed RRH scenario, the SINDR of a UE is predominantly influenced by its nearest RRH, and consequently, the dithering level at the RRH mostly affects its closest UE. Therefore, the proposed criteria to tune the RRH dithering levels is based on the nearest UE. To find an initial compromise set of RRH dithering levels based on the nearest UE at each RRH, we define $\delta_{b}^{\max} \triangleq \max_{k \in \setK} \delta_{b,k}$ to denote the maximum path gain (minimum path loss) between RRH \(b\) and its closest UE in $\setK$. Furthermore, if the nearest UE of an RRH is closer than the UEs near other RRHs, that RRH saturates more quickly, necessitating the addition of dithering earlier than other RRHs. As a result, $\frac{\sigma_b}{\sqrt{\delta_{b}^{\max}}}$ becomes more similar among RRHs, improving the performance of the joint reception.
Consequently, the RRH dithering levels $\{\sigma_{b}\}$ are updated using a single parameter $\theta$ 
as 
\begin{align}\label{eq:sigma_b_ud}
\sigma_{b} = \sigma_{\min} \max \big(1, \sqrt{{\delta_{b}^{\max}}} \ \theta\big)
\end{align}
where the search variable $\theta$ lies within the range of $[1/ \max_{\bar b \in \setB}\sqrt{\delta_{b}^{\max}} , 1/\min_{\bar b \in \setB} \sqrt{\delta_{b}^{\max}} ]$. 
Clearly, if $\theta = 1/ \max_{\bar b \in \setB}\sqrt{\delta_{b}^{\max}}$, dithering is set to $\sigma_{b}=\sigma_{\min} \ \forall \ b$. Conversely, if $\theta = 1/ \min_{\bar b \in \setB}\sqrt{\delta_{b}^{\max}}$, dithering is increased at all the RRHs except the one farthest from any UE in $\setK$.  In general, provided that there are no limits on the UE transmit powers, increasing the RRH dithering improves the feasible range of SINDR values. However, for a fixed feasible SINDR target, the min-power may exhibit a unimodal behavior with respect to the value of $\theta$. Therefore, we use a line search method to find the optimal $\theta$ value corresponding to the minimum sumpower. Specifically, we employ a ternary search method, evaluating the UE transmit power for two different values of $\theta$ within its defined limits~\cite{mis}. 
These two points are selected to divide the entire range of $\theta$ into three equal intervals. Subsequently, we compute the sum power at these two values of $\theta$ and adjust the range of $\theta$ values in the next iteration, such that the value of $\theta$ corresponding to the minimum sum power lies in the middle of the updated range, as detailed in Algorithm~\ref{alg:pwr_noise_minPwr}.

 \begin{figure}[t!]
\begin{minipage}[t]{0.4\textwidth}
\begin{algorithm}[H]
\footnotesize
\textbf{Data:} $ \{\hat \h_k \}$, $ \{\C_{\tilde \h_k} \}$, $\{\gamma_k^{\tar} \}$, $\rho^{\max}$, $\epsilon_{\sigma} > 0$, $\epsilon_{\rho} > 0$. \\
\textbf{Initialization:} Iteration $i=0$,  $\theta^{\low} =  1/ \max_{ b \in \setB} \sqrt{\delta_{b}^{\max}}$, \\  $\theta^{\high}  =  1 /\min_{b \in \setB} \sqrt{\delta_{b}^{\max}}$. \\
\%Finding the coarse RRH dithering levels.

\While{$(1)$} 
{
\%Evaluate the min-power at two points.

 \For {$l=1:2$} 
    { 
    \begin{itemize}
     \item $\theta^{(l)} \triangleq  {\big((3-l)\theta^{\low} +l \theta^{\high} \big)}/{3}$ 
        \item Compute $\{\sigma_b\}$ using \eqref{eq:sigma_b_ud}.
        \item Compute $\{\rho_k^{(l)}\}$ using \eqref{eq:gdUp_pwr} or Algorithm~\ref{alg:pwrScal}. 
    \end{itemize}
     \If{$\min_{k \in \setK} (\hat \gamma_k - \gamma^{\tar}_k)\le 0$} 
     {
     \begin{itemize}
         \item $\rho_k^{(l)} = \rho^{\max}/(\theta^{(l)}),~\forall k \in \setK.$
     \end{itemize}
     }
    }
    \eIf{$\sum_{k \in \setK} \rho^{(1)}_k > \sum_{k \in \setK} \rho^{(2)}_k$} 
     {
     \begin{itemize}
        \item $\theta^{\low} =  {\big(2\theta^{\low} + \theta^{\high} \big)}/{3}$.
         \end{itemize}
 }{   
     \begin{itemize}
        \item  $\theta^{\high} =   {\big(\theta^{\low} + 2\theta^{\high} \big)}/{3}.$
    \end{itemize}
     }
        \If{$(\theta^{\low} - \theta^{\high}) \le \epsilon_{\sigma}$}
   {
   \textbf{break} 
   }
  }
\% Fine-tuning the RRH dithering and UE transmit powers.

   \While{ $\big(\sum_{k \in \setK} \rho^{(i)}_k - \sum_{k \in \setK} \rho^{(i-1)}_k \big) \le \epsilon_{\rho}$}  
        {
          \begin{itemize}
          \item $i=i+1$
               \item Compute $\{\sigma_b\}$ using \eqref{eq:gdUp_noise}.
        \item Compute $\{\rho_k^{(i)}\}$ using \eqref{eq:gdUp_pwr} or \eqref{eq:pwrScal}. 
          \end{itemize}
        }
\caption{Optimization of the UE transmit powers and RRH dithering levels with the min-power objective} \label{alg:pwr_noise_minPwr}
\end{algorithm}
\end{minipage}
\vspace{-5mm}
\end{figure}

\smallskip

\begin{figure*}
\setcounter{equation}{33}
\begin{align}
\label{eq:p3gradLag_bmrc} \frac{\partial}{\partial \sigma_b} \mathcal{L}_{\eqref{eq:p1}}^{\bmrc}\big(\{\sigma_b, \mu_k^{\bmrc}, \upsilon_b \}\big) & = -\sum_{\bar k \in \setK}\mu_{\bar k}^{\bmrc} \bigg( 4\rho_k (\gamma_{\bar k}^{\tar}+1)   \hat \h_{\bar k}^{\herm} \hat \A \hat \A  \hat \h_{\bar k} \Re \bigg[ \hat \h_{\bar  k}^{\herm} \Big(\frac{\partial}{\partial \sigma_{b}} \hat \A \Big) \hat \A \hat \h_{\bar  k}\bigg] \nonumber \\  
 & \phantom{=} \ -  2\gamma_k^{\tar} \Re \bigg[\hat \h_{\bar  k}^{\herm} \Big(\frac{\partial}{\partial \sigma_{b}}\hat \A \Big) \C_{\hat \r} \hat \A  \hat \h_{\bar  k} \bigg] - \gamma_k^{\tar}  \hat \h_{\bar k}^{\herm} \hat \A  \Big( \frac{\partial}{\partial \sigma_{b}}  \C_{\hat \r}  \Big) \hat \A  \hat \h_{ \bar k} \bigg) - \upsilon_b
\end{align}
\hrule
\begin{align}
\label{eq:p3gradLag_bmmse} \frac{\partial}{\partial \sigma_b} \mathcal{L}_{\eqref{eq:p1}}^{\bmmse}\big(\{\sigma_b, \mu_k^{\bmmse}, \upsilon_b\}\big)& = - \sum_{ \bar k \in \setK} \mu_{\bar k}^{\bmmse} (\gamma_{\bar k}^{\tar} + 1)\rho_{\bar k}  \bigg( 2 \Re \bigg[\hat \h_{\bar  k}^{\herm} \Big(\frac{\partial}{\partial \sigma_b}\hat \A \Big) \C_{\hat \r}^{-1} \hat \A \hat \h_{\bar  k} \bigg] - \hat \h_{\bar k}^{\herm} \hat \A \C_{\hat \r}^{-1} \Big( \frac{\partial}{\partial \sigma_{b}}  \C_{\hat \r}  \Big) \C_{\hat \r}^{-1} \hat \A  \hat \h_{ \bar k}\bigg) \! - \!\upsilon_b
\end{align}
\setcounter{equation}{31}
\hrule
\end{figure*}

After optimizing the UE transmit powers in \eqref{eq:p1} with the coarse dithering levels from \eqref{eq:sigma_b_ud}, both the UE transmit powers and RRH dithering levels can be fine-tuned using the gradient method in an alternating manner. For fixed RRH dithering levels, the UE transmit powers are optimized with the gradient method described in Section~\ref{sec:min_pwr_tp}. For fixed UE transmit powers, the coarse RRH dithering levels are fine-tuned using the gradient method. 
To do this, for a fixed $\{\rho_k\}$, we construct the Lagrangian of \eqref{eq:p1} for a given receiver as
\begin{align}
\label{eq:p3Lag_bmrc}
\mathcal{L}_{\eqref{eq:p1}}^{\rx}\big(\{\sigma_b, \mu_k^{\rx}, \upsilon_b\}\big) & \triangleq \sum_{k \in \setK} (\rho_k -\mu_k^{\rx} \xi_k^{\rx}) \!-\! \sum_{b \in \setB} \upsilon_b (\sigma_b - \sigma_{\min}), 
\end{align}
where $\upsilon_b \ge 0$ is the Lagrangian dual variable corresponding to the minimum noise (dithering) level constraint in \eqref{eq:p1}. The dithering level $\sigma_b$ is then updated iteratively in the direction of the negative gradient of the Lagrangian as 
\begin{equation}
\begin{aligned}
\label{eq:gdUp_noise}
\hspace{-2mm}\sigma_b^{(i)} = \max\big(\sigma_{\min},\sigma_b^{(i-1)} - \zeta_{\npwr} \frac{\partial}{\partial \sigma_b} \mathcal{L}_{\eqref{eq:p1}}^{\rx}\big(\{\sigma_b, \mu_k^{\rx}, \upsilon_b\}\big) \big),
\end{aligned}
\end{equation}
where $\zeta_{\npwr} \ge 0$ is the step size and $i$ is the iteration number. For the initial iteration, the values of ${\sigma_b}$ are obtained from \eqref{eq:sigma_b_ud}. The gradient of the Lagrangian with respect to $\sigma_b$ is given in \eqref{eq:p3gradLag_bmrc} and \eqref{eq:p3gradLag_bmmse} at the top of the next page with the BMRC and BMMSE receivers, respectively, whereas the partial derivative of $\hat \A$ and $\C_{\hat \r}$ with respect to $\sigma_b$ are provided in Appendix~\ref{app:grd_noise_min}. \setcounter{equation}{35} The dual variable $ \mu_k^{\rx}$ is updated as in \eqref{eq:mu_up_bmrc}. Similarly, the dual variable $\upsilon_b$ is updated in the same way for both the BMRC and BMMSE receivers: 
\begin{align} \label{eq:zeta_ud}
\upsilon_b^{(i)} = \max \big(0, \upsilon_b^{(i-1)} - \kappa (\sigma_b -\sigma_{\min}) \big),
\end{align}
where $\kappa \ge 0$ is the steps size. For the initial iteration, $\{\upsilon_b\}$ are initialized with positive random values. 

The fine-tuning of the UE transmit powers and RRH dithering levels involves a single gradient update in each iteration. This process continues until there is no significant improvement in the sum power.
The full procedure to optimize the UE transmit powers and the RRH dithering levels with the min-power objective is summarized in Algorithm~\ref{alg:pwr_noise_minPwr}.

\section{Uplink Transmit Power Optimization Based on Max-Min-SINDR}\label{sec:max_min_SINDR}

In this section, we consider the maximization of the minimum SINDR across all the UEs subject to UE-specific UL transmit power constraints $\rho_{\ue}$. Similar to Section~\ref{sec:min_pwr}, both the UE transmit powers and RRH dithering levels are treated as optimization variables. Thus, the epigraph form of the max-min-SINDR optimization problem is given by
\begin{align} \label{eq:p5}
\begin{array}{cl}
\underset{\{\rho_{k}, \sigma_b \}, \gamma}{\maximize} &   \gamma\\
\mathrm{s.t.} &  \hat \gamma_k \ge \gamma, \quad \forall k \in \setK \\ 
& \rho_k \le \rho_{\ue}, \quad \forall k \in \setK \\
 & \sigma_{b} \ge \sigma_{\min}. \quad \forall b \in \setB,
\end{array}
\end{align}
where $\gamma$ is the optimization variable representing the minimum rate across all the UEs. The UE-specific SINDR value $\hat \gamma_k$ can be obtained using either BMRC or BMMSE receivers given in \eqref{eq:bmrc_sinr_k} or \eqref{eq:bmmse_sinr_k}, respectively. Problem \eqref{eq:p5} can be solved using min-power objective as in \eqref{eq:p1} by fixing $\gamma$ and employing the bisection method to determine the optimal value of $\gamma$. However, in the subsequent discussion, we explore alternative techniques to solve \eqref{eq:p5} without relying on the bisection search for $\gamma$. As \eqref{eq:p5} is also not convex with respect to $\rho_k$ and $\sigma_b$, we first discuss the optimization of the UE transmit powers for fixed RRH dithering levels as in Section~\ref{sec:min_pwr}. Then, we proceed to discuss the optimization of the RRH dithering levels.

\subsection{Optimization of the UE Transmit Powers}

\begin{figure}[t!]
\begin{minipage}[t]{0.9\linewidth}
\begin{algorithm}[H]
\footnotesize
\textbf{Data:} $\{\hat \h_k\}$, $\{\C_{\tilde \h_k}\}$, $\rho_{\ue}$, $\beta$. \\  
\textbf{Initialization:} Iteration~$i=0$, $\{\rho_k^{(0)}\}$. 
\begin{itemize}
    \item Compute $\{\hat \sinr_k^{(0)}\}$ as in \eqref{eq:bmrc_sinr_k} or \eqref{eq:bmmse_sinr_k}.
    \item $\gamma^{(0)} \triangleq \min_{k \in \setK} \hat\sinr_k^{(0)}$.
\end{itemize}
\While{ $\max_{k \in \setK} {\rho_k^{(i)}} \le \rho_{\ue}$}
{
   \begin{itemize}
       \item $\bar k = \argmin_{k \in \setK} \hat \gamma_k^{(i)}$.
       \item $i=i+1$.
       \item $\rho^{(i)}_{\bar k} = \beta \rho^{(i-1)}_{\bar k}$, and  $\rho^{(i)}_{k} = \rho^{(i-1)}_{k}, \quad \forall k \in \setK \setminus \{ \bar k\}.$
       \item Compute $\{\hat \sinr_k^{(i)}\}$ as in \eqref{eq:bmrc_sinr_k} or \eqref{eq:bmmse_sinr_k}.
       \item $\gamma^{(i)} \triangleq \min_{k \in \setK} \hat \sinr_k^{(i)}$.
   \end{itemize} 
   \If{ $\hat \gamma_{\bar k}^{(i)} \! < \! \hat \gamma_{\bar k}^{(i-1)} \textnormal{\textbf{or}} \ \max_{k \in \setK} \hat \sinr_k^{(i)} \! < \! \max\limits_{j=\{0, \ldots, i\}}\gamma^{(j)}\ $}
   {
   \textbf{break}
   }
}
\begin{itemize}
    \item The optimal SINDR $\sinr = \max_{j=\{0, \ldots, i\}}\gamma^{(j)}$.
\end{itemize}
\caption{Optimization of the UE transmit powers with the max-min-SINDR objective via the BCD method} \label{alg:BCDpwrSea}
\end{algorithm}
\end{minipage}
\end{figure}
For a fixed set of RRH dithering levels, we may choose to optimize the UE transmit powers of problem \eqref{eq:p5} using the gradient method similar to that in Section~\ref{sec:min_pwr}. However, the gradient update of \eqref{eq:p5} becomes more complex and slow due to the extra variable $\gamma$. Therefore, we employ a BCD method akin to the fixed-point power update, while considering the impact QD on the SINDR of UEs as in Section~\ref{sec:min_pwr}.

\smallskip

\textit{\textbf{BCD method.}} The BCD method for the UE transmit power optimization, described in Section~\ref{sec:min_pwr} for the min-power objective, can be adapted for the max-min SINDR objective. The target SINDR objective in \eqref{eq:p5} is the same for all the UEs. Thus, the user selection in \eqref{eq:ue_idx} simplifies to finding a UE with the minimum SINDR, i.e., $k = \argmin_{\bar k \in \setK} {\hat \sinr_{\bar k}^{(i-1)}}$. Since the target SINDR is a variable and not known, only the fixed scaling factor $\alpha > 1$ in \eqref{eq:pwrScal} is used to update the transmit power of the selected UE, while keeping the transmit powers of the other UEs unchanged. 
Consequently, the minimum SINDR is increased in every iteration as long as the SINDR values remain in the AWGN-dominated region for an appropriate choice of $\beta$. 
The BCD method is terminated if any UE reaches its maximum power or if the SINDR of any UE decreases despite increasing its transmit power,  while keeping the other UE transmit powers fixed, (i.e., $\gamma_k^{(i)}<\gamma_k^{(i-1)}$ if $\rho^{(i)}>\rho^{(i-1)}$) as defined in Definition~\ref{def:nonmonQD}.
Additionally, the process is terminated if the SINDR of all the UEs decreases with increasing their transmit powers (i.e., $\max_{k \in \mathcal{K}} \gamma_k(\boldsymbol{\rho}) < \gamma(\boldsymbol{\rho'})$ if  $\boldsymbol{\rho} \succeq \boldsymbol{\rho'}$),  
as stated in Lemma~\ref{thm:qdsc}. 
At the end of the BCD method, the maximum of minimum SINDRs over the iterations is considered as the max-min SINDR, i.e., $\sinr = \max_{j=\{0, \ldots, i\}}\gamma^{(j)}$. The full procedure to optimize the UE transmit powers with the max-min-SINDR objective via the BCD method is summarized in Algorithm~\ref{alg:BCDpwrSea}.


\begin{lemma}\label{thm:qdsc}
The transmit power of a UE is in the QD-dominated region of the non-monotonic SINDR function if there exists a UE transmit power vector $\boldsymbol{\rho} \succeq \boldsymbol{\rho'}$ such that $\max_{k \in \mathcal{K}} \gamma_k(\boldsymbol{\rho}) < \gamma(\boldsymbol{\rho'})$, where $\gamma(\boldsymbol{\rho'}) = \min_{k \in \mathcal{K}} \gamma_k(\boldsymbol{\rho'})$.
\end{lemma}

\begin{IEEEproof}
The condition $\max_{k \in \mathcal{K}} \gamma_k(\boldsymbol{\rho}) < \gamma(\boldsymbol{\rho'})$ can be rewritten as $\gamma_k(\boldsymbol{\rho}) < \gamma_k(\boldsymbol{\rho'}),~\forall k \in \mathcal{K}$. Then, the rest of the proof follows similar steps to the proof of Theorem~\ref{thm:qdscMinpwr}.
\end{IEEEproof}

\smallskip



\subsection{Optimization of the RRH Dithering Levels}

To find the jointly (locally) optimal solution for \eqref{eq:p5}, the UE transmit powers need to be optimized for different values of RRH dithering levels.
To avoid the complexity of searching for all the possible RRH dithering levels, we choose to employ the same heuristic technique as discussed in Section~\ref{sec:min_pwr} to obtain coarse initial RRH dithering values of $\{\sigma_b\}$ from a single variable $\theta$ as in \eqref{eq:sigma_b_ud}. Considering the limitation on the UE transmit powers, the SINDR is a unimodal function with respect to $\theta$ or RRH dithering levels. Therefore, we can employ the line search method similar to the min-power design (as in Algorithm~\ref{alg:pwr_noise_minPwr}) to find the suitable value of theta $\theta$. However, in this design, the range of $\theta$ values is updated based on maximizing the minimum SINDR across all the UEs in each iteration, as detailed in Algorithm~\ref{alg:pwr_noise_maxmin}.

As in Section~\ref{sec:min_pwr}, after the coarse optimization of the UE transmit powers and RRH dithering levels. We can further fine-tune them via alternating optimization. For fixed values of $\{\sigma_b\}$, the UE transmit powers can be optimized using the BCD method (in Algorithm~\ref{alg:BCDpwrSea}). For a fixed UE transmit power, we can further adjust the RRH dithering levels $\{\sigma_b\}$ using the gradient method to improve the value of $\gamma$. To do this, for a fixed $\{\rho_k\}$, we construct the Lagrangian of \eqref{eq:p5} for a given receiver as  
\begin{align}
\label{eq:p7Lag_bmrc}
\mathcal{L}^{\rx}_{\eqref{eq:p5}}\big(\{\sigma_b, \gamma, \eta_k^{\rx}, \upsilon_b\}\big) \! & \triangleq \! \gamma \!- \! \sum_{k \in \setK}  \! \eta_k^{\rx} \chi_k^{\rx} \! - \! \sum_{b \in \setB} \! \upsilon_b (\sigma_b \! - \! \sigma_{\min}),
\end{align}
where $\chi_k^{\rx}$ represents the SINDR constraint in \eqref{eq:p5}, obtained by replacing $\gamma_k^{\tar}$ with $\gamma$ in $\eqref{eq:sindr_con_bmrc}$ and $\eqref{eq:sindr_con_bmmse}$ corresponding to the BMRC and BMMSE receivers, respectively. $\eta_k^{\rx} \ge 0 $ are the Lagrangian dual variable associated with the SINDR constraint in \eqref{eq:p5}. Lastly, $\upsilon_b \ge 0$ is the Lagrangian dual variable corresponding to the minimum dithering (noise) level in \eqref{eq:p5}. Subsequently, for a given receiver, the dithering level at RRH $b$ is iteratively updated in the direction of the negative gradient of the Lagrangian as follows:
\begin{equation}
\sigma_b^{(i)} = \max\Big(\sigma_{\min}, \sigma_b^{(i-1)} - \zeta_{\npwr} \frac{\partial}{\partial \sigma_b} \mathcal{L}_{\eqref{eq:p5}}^{\rx}\big(\{\sigma_b, \gamma, \eta_k^{\rx}, \upsilon_b\}\big) \Big).
\label{eq:gdUp_noise_maxmin}
\end{equation}
The gradient of the Lagrangian can be obtained by replacing $\gamma_k^{\tar}$ with $\gamma,~\forall k \in \setK$, in \eqref{eq:p3gradLag_bmrc} and \eqref{eq:p3gradLag_bmmse} with the BMRC and BMMSE receivers, respectively. For the initial iteration, the values of $\sigma_b$ are obtained from \eqref{eq:sigma_b_ud}. The dual variable $\eta_k^{\rx}$ is then updated as in \eqref{eq:mu_up_bmrc}, replacing $\gamma_k^{\tar}$ with $\gamma,~\forall k \in \setK$, and the dual variable $\upsilon_b$ is updated as in \eqref{eq:zeta_ud}. To simplify the computation of $\gamma$, let us define $a_k^{\bmrc} \triangleq \rho_k \left| \hat \h_k ^{\herm} \hat \A \hat \A \hat \h_k \right|^2$, $b_k^{\bmrc} \triangleq \hat \h_k ^{\herm} \hat \A \C{\hat \r} \hat \A \hat \h_k $, $a_k^{\bmmse} \triangleq  \rho_k \hat \h_k^{\herm} \hat \A \C{\hat \r}^{-1} \hat \A \hat \h_k$, $b_k^{\bmrc} \triangleq 1$. With these definitions, for a given receiver, the complementary slackness condition of the SINDR constraint in \eqref{eq:p5}, for all the UEs is written as~\cite{Mah21}
\begin{align}
    \sum_{k \in \setK} \eta_k^{\rx} \big( (\gamma+1)a_k^{\rx} - \gamma b_k^{\rx}\big) =0.
\end{align}
As a result, the value of $\gamma$ can be updated as
\begin{align}
  \gamma = \frac{\sum_{k \in \setK} \eta_k^{\rx} a_k^{\rx}}{ \sum_{k \in \setK} \eta_k^{\rx}(b_k^{\rx} -a_k^{\rx})}.
 \end{align}
However, at the end of the iterative process, we have $\gamma \triangleq \min_{k \in \setK} \hat \sinr_k$. For the fine-tuning a single gradient update of the RRH dithering levels is used for every update of the UE transmit powers. Finally, the fine-tuning of the UE transmit powers and RRH dithering levels terminates if there is no significant improvement in the minimum SINDR. The full procedure to optimize the UE transmit powers and the RRH dithering levels with the max-min-SINDR objective is summarized in Algorithm~\ref{alg:pwr_noise_maxmin}.

\begin{figure}[t!]
\begin{minipage}[t]{0.86\linewidth}
\begin{algorithm}[H]
\footnotesize
\textbf{Data:} $\{\hat \h_k\}$, $\{\C_{\tilde \h_k}\}$, $\epsilon_{\sigma} > 0$, $\epsilon_{\gamma} > 0$. \\
\textbf{Initialization:} Iteration $i=0$,  $\theta^{\low} =  1/ \max_{ b \in \setB} \sqrt{\delta_{b}^{\max}}$, \\  $\theta^{\high}  =  1 /\min_{ b \in \setB} \sqrt{\delta_{b}^{\max}}$.\\
\%Coarse RRH dithering level optimization.

\While{$(1)$}
{
\%Evaluate the max-min-SINDR at two points.
 
 \For {$l=1:2$} 
    { 
    \begin{itemize}
     \item $\theta^{(l)} \triangleq  {\big((3-l)\theta^{\low} +l \theta^{\high} \big)}/{3}$. 
        \item Compute $\{\sigma_b\}$ using \eqref{eq:sigma_b_ud}.
        \item Compute $\gamma^{(l)}$ using Algorithm~\ref{alg:BCDpwrSea}.
    \end{itemize}
    }
    \eIf{$\gamma^{(1)} > \gamma^{(2)}$} 
     {
     \begin{itemize}
         \item $\theta^{\low} = {\big(2\theta^{\low} + \theta^{\high} \big)}/{3}$.
         \end{itemize}
 }{   
     \begin{itemize}
        \item  $\theta^{\high} =  {\big(\theta^{\low} + 2\theta^{\high} \big)}/{3}.$
    \end{itemize}
     }

        \If{$(\theta^{\low} - \theta^{\high}) \le \epsilon_{\sigma}$}
   {
   \textbf{break} 
   }
  }

\% Fine-tuning the RRH dithering and UE transmit powers.

   \While{ $(\gamma^{(i)} - \gamma^{(i-1)}) \le \epsilon_{\sigma}$}  
        {
          \begin{itemize}
          \item $i=i+1$
               \item Compute $\{\sigma_b\}$ using \eqref{eq:gdUp_noise_maxmin}.
        \item Compute $\{\rho_k^{(i)}\}$  using Algorithm~\ref{alg:BCDpwrSea}.
          \end{itemize}
        }
\caption{Optimization of the UE transmit powers and RRH dithering levels with the max-min-SINDR objective} \label{alg:pwr_noise_maxmin}
\end{algorithm}
\end{minipage}
\vspace{-3mm}
\end{figure}

\section{Numerical Results} \label{sec:NUM}

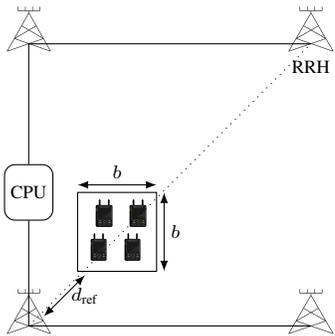
\begin{figure}[t!]
\begin{center}
\begin{tikzpicture}

\pgfdeclareimage{UE}{UE}
\pgfdeclareimage{BS}{BS}

\usetikzlibrary{calc}

\newdimen\d
\d=1.5cm;

\tikzstyle{block}=[rectangle, draw, text centered, rounded corners, minimum height=3em]

\scriptsize

\draw node[block] (CPU) {CPU};

\draw node[below=1\d of CPU, scale=0.005, node distance=1\d] (BS_1) {\pgfbox[center,bottom]{\pgfuseimage{BS}}};
\draw node[right=2.5\d of BS_1, scale=0.005, node distance=2\d] (BS_2) {\pgfbox[center,bottom]{\pgfuseimage{BS}}};
\draw node[above=1\d of CPU, scale=0.005, node distance=1\d] (BS_3) {\pgfbox[center,bottom]{\pgfuseimage{BS}}};
\draw node[right=2.5\d of BS_3, scale=0.005, node distance=2\d] (BS_4) {\pgfbox[center,bottom]{\pgfuseimage{BS}}};

\draw node[below right=0.35\d and 0.4\d of CPU, scale=0.005, node distance=0.5cm] (UE_1) {\pgfbox[center,bottom]{\pgfuseimage{UE}}};
\draw node[below right=0.35\d and 0.7\d of CPU, scale=0.005, node distance=0.5cm] (UE_2) {\pgfbox[center,bottom]{\pgfuseimage{UE}}};
\draw node[below right=0.05\d and 0.45\d of CPU, scale=0.005, node distance=0.5cm] (UE_3) {\pgfbox[center,bottom]{\pgfuseimage{UE}}};
\draw node[below right=0.05\d and 0.75\d of CPU, scale=0.005, node distance=0.5cm] (UE_4) {\pgfbox[center,bottom]{\pgfuseimage{UE}}};

\draw [-] ($(BS_1.north)+(0mm,1mm)$) -- ($(BS_2.north)+(0mm,1mm)$) node {};
\draw [-] ($(BS_3.north)+(0mm,1mm)$) -- ($(BS_4.north)+(0mm,1mm)$) node {};
\draw [-] (CPU) -- ($(BS_1.north)+(0mm,1mm)$) node {};
\draw [-] (CPU) -- ($(BS_3.north)+(0mm,1mm)$) node {};
\draw [dotted] ($(BS_1.north)+(0mm,1mm)$) -- ($(BS_4.north)+(0mm,1mm)$) node {};

\begin{scope}[>=latex]
\draw (0.65,-1.05) rectangle (1.7,0) {};
\draw [<->] (0.65,0.1) -- (1.7,0.1) node[midway,above]{$b$};
\draw [<->] (1.8,0) -- (1.8,-1.05) node[midway,right]{$b$};
\draw [<->] (0.2,-1.65) -- (0.75,-1.1) node[midway,right] {$d_{\textrm{ref}}$};
\end{scope}[>=latex]
\draw node[below=0.025*\d of BS_4] {RRH};
\end{tikzpicture}
\end{center}
\caption{Distributed massive MIMO system considered in the numerical analysis.}
\label{fig:dMIMO_setup}
\vspace{-3mm}
\end{figure}

To evaluate the performance of the distributed massive MIMO system with 1-bit ADCs, we consider $B=4$ RRHs, each equipped with $M=256$ antennas, positioned at the corners of a square grid with distance to the closest neighboring RRH of $100$~m, and the height of each RRH is $5$~m. All the RRHs are connected to the central processing unit (CPU). Unless otherwise stated, we consider $K=4$ single-antenna UEs placed at the corners of a square cluster of size $b \times b$~m$^2$, as depicted in Fig.~\ref{fig:dMIMO_setup}. The minimum distance between the closest RRH and the UE cluster is denoted by $d_{\textrm{ref}}$. The AWGN power at the RRH is set to $\sigma^2_{\min}=-95$~dBm and additional dithering (i.i.d noise) can be added if necessary to enhance the performance. Details of the channel model are provided in Section~\ref{sec:SINDRChar}. We employ orthogonal time resources for the channel estimation of each UE with a pilot length of $128$. Additionally, this estimation utilizes the maximum UE transmit powers and adjusts the RRH dithering levels based on the pathloss or the received signal power. Finally, the UE transmit power for the min-power design is optimized using the \textit{Gradient} and \textit{BCD} methods, while for the max-min-SINDR, it is optimized using the \textit{BCD} method, as discussed in Sections~\ref{sec:min_pwr} and~\ref{sec:max_min_SINDR}, respectively.

          \begin{figure}[t]
          \centering
            \input{FigJor/Fig5.tex}
            \vspace{-1mm}
            \caption{Min-power versus target SINDR with dithering.} \label{fig:minPwrVsSINDR}
            \vspace{-3mm}
        \end{figure}
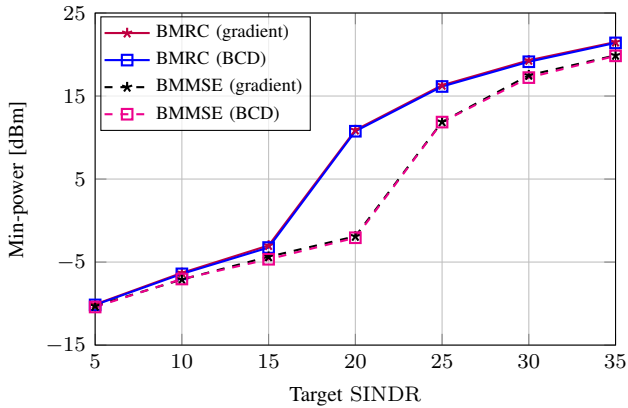
Fig.~\ref{fig:minPwrVsSINDR} illustrates the relationship between the min-power and the target SINDR with both the BMRC and BMMSE receivers with $d_{\text{ref}}=0$ and $b =10$~m. To achieve higher SINDR targets through joint reception from multiple RRHs, dithering is introduced at all the RRHs except the farthest one. The \textit{Gradient} and \textit{BCD} methods exhibit comparable performance. The computational complexity of the \textit{Gradient} method in each iteration with BMRC and BMMSE receivers, determined by \eqref{eq:gradLag_bmrc} and  \eqref{eq:gradLag_bmmse}, respectively, is higher compared to the \textit{BCD} method. Convergence is also influenced by the step sizes, denoted by $\zeta_{\rho}$ (in \eqref{eq:gdUp_pwr}) and $\nu$ (in \eqref{eq:mu_up_bmrc}). Furthermore, in both the \textit{Gradient} and \textit{BCD} methods, the computational complexity with the BMMSE receiver exceeds that of the BMRC receiver, as the former involves matrix inversion operations. However, to achieve the same target SINDR, the BMRC receiver requires slightly more transmit power compared to the BMMSE receiver at high SINDR, as the latter effectively reduces the impact of the QD and interference between the UEs. This reduction in QD and interference with the BMMSE receiver also facilitates achieving a higher target SINDR, with or without adding dithering at the RRHs, for a given min-power. Dithering at the RRH is not required to achieve target SINDRs of $15$ or less for BMRC receivers and $20$ or less for BMMSE receivers. However, the joint reception from multiple RRHs and dithering at the closest RRHs are necessary to achieve the target SINDRs of $20$ and $25$ or more with BMRC and BMMSE receivers, respectively, while minimizing the sum of UE transmit powers. 
Consequently, a sharp increase in the min-power is observed, which is required to combat the pathloss to the faraway RRHs, resulting in a significant power difference between the BMRC and BMMSE receivers at a target SINDR of $20$.

        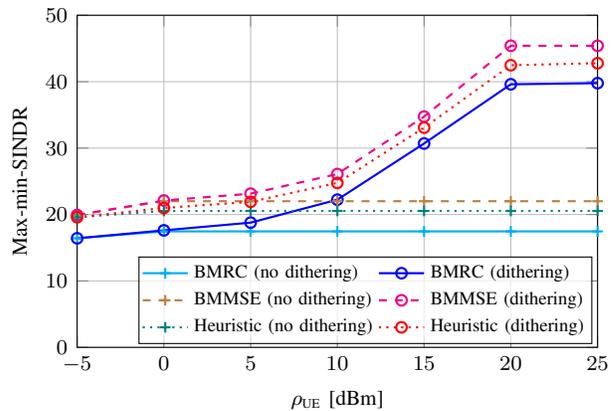
\begin{figure}[t]
        \centering
            \input{FigJor/fig6.tex}
             \vspace{-1mm}
            \caption{Max-min-SINDR versus maximum UE transmit power.} \label{fig:PwrVsmmSINDR}
            \vspace{-3mm}
        \end{figure}

          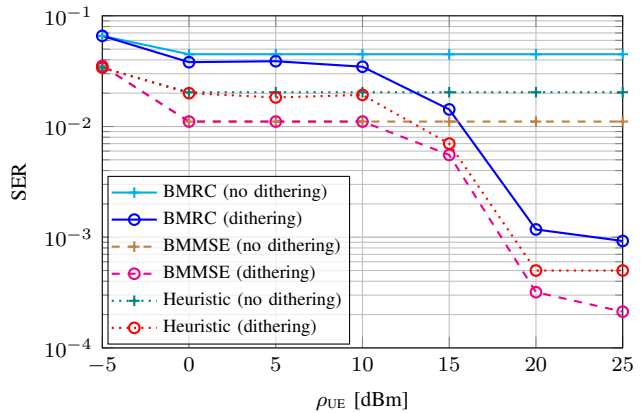
\begin{figure}[t]
          \centering
            \input{FigJor/fig7.tex}
             \vspace{-1mm}
            \caption{SER versus maximum UE transmit power with 16-QAM data symbols.} \label{fig:PwrVsSER}
            \vspace{-3mm}
        \end{figure}
 
Fig.~\ref{fig:PwrVsmmSINDR} illustrates the achievable max-min-SINDR for a given maximum UE transmit power $\rho_{\ue}$ with $d_{\text{ref}}=0$ and $b =10$~m. As expected, the BMMSE receiver outperforms the BMRC receiver in both scenarios, with and without dithering. Without dithering, the maximum achievable SINDR is around $17$ for the BMRC receiver and $20$ for the BMMSE receiver, respectively, while with joint reception from multiple RRHs and dithering, the maximum achievable SINDR increases to $40$ and $42$ for BMRC and BMMSE receivers, respectively. The impact of the joint reception from multiple RRHs with dithering is notable when the UE transmit power exceeds $5$~dBm, as it implies that the SINDR is in the QD-dominated region with respect to the closest RRH and the UE has enough power to reach the other RRHs. To reduce the computational complexity of UE transmit power and dithering optimizations, we propose using the BMRC receiver for solving the max-min-SINDR objective and then switching to the BMMSE receiver for the data reception. This method is referred to as \textit{Heuristic}. While the performance of \textit{Heuristic} is inferior to that of the BMMSE receiver, it is better than the performance of the case where BMRC receiver is used also for data detection, with and without dithering. Considering the same setup as in Fig.~\ref{fig:PwrVsmmSINDR}, we plot the maximum SER of a UE in Fig.~\ref{fig:PwrVsSER} with 16-QAM data symbols. The SER exhibits a similar behavior to the max-min-SINDR. The lowest SER without dithering is about $10^{-2}$, whereas with joint reception and dithering, it is around $2\times 10^{-4}$ at the UE transmit power of $25$~dBm. Additionally, The performance of the \textit{Heuristic} method surpasses that of the BMRC receiver.

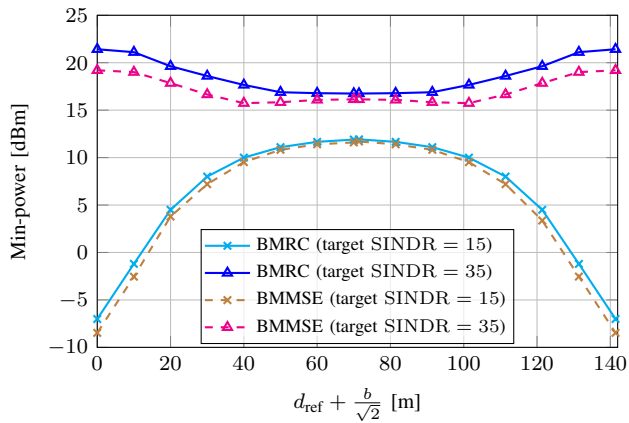
\begin{figure}[t]
\centering
            \input{FigJor/Fig8.tex}
             \vspace{-1mm}
            \caption{Min-power versus distance to the reference RRH.} \label{fig:DisvsPwr}
            \vspace{-4mm}
        \end{figure}

Fig.~\ref{fig:DisvsPwr} illustrates the minimum power required to achieve a given target SINDR as a function of the distance between the center of the UE cluster and the bottom-left RRH (also referred to as the \textit{reference RRH}), while the center of the UE cluster moves along the dotted lines from the bottom-left RRH to the top-right RRH, as shown in Fig.~\ref{fig:dMIMO_setup} with $b = 10$~m. If the target SINDR is achieved by a single RRH, such as a SINDR of $15$, the required transmit power increases as the center of the UE cluster moves away from the RRH, as expected in the unquantized system. Therefore, the maximum min-power is required when the center of the UE cluster is located in the middle of the bottom-left and the top-right RRHs. However, if the target SINDR requires joint reception from all the RRHs with dithering, as in the case of a SINDR of $35$, the min-power required to achieve the target SINDR is maximum when the center of the UE cluster is closer to one of the RRHs and minimum when it is equidistant from all the RRHs. This is because when the UEs are close to an RRH, they require more transmit power to reach the far-away RRHs for joint reception with dithering. This result differs from the unquantized system. As expected, the min-power required at any distance with the BMMSE receiver is less than with the BMRC receiver in both scenarios. However, at high SINDR, the performance of the BMMSE receiver is much better than the BMRC receiver compared to the low SINDR.

               \begin{figure}[t]
               \centering
            \input{FigJor/fig9.tex}
             \vspace{-1mm}
            \caption{Max-min-SINDR versus distance to the reference RRH.} \label{fig:DisvsmmSINDR}
            \vspace{-4mm}
        \end{figure}
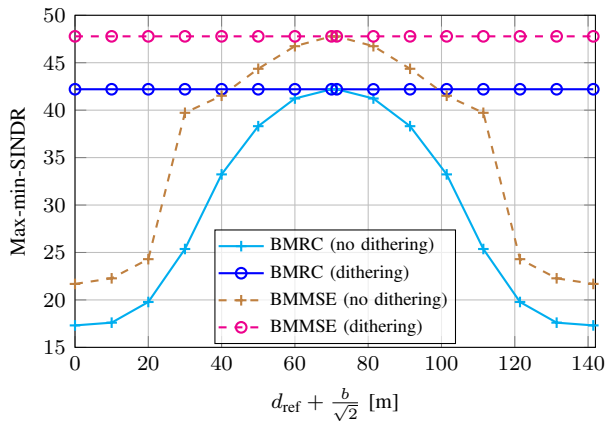

In Fig.~\ref{fig:DisvsmmSINDR}, we 
characterize the max-min-SINDR as a function of the distance to the center of the UE cluster to the reference RRH with $b = 10$~m, considering a maximum UE transmit power of $25$~dBm. When dithering is not added, the SINDR is primarily determined by the closest RRH to the UE cluster, due to the SINDR transitioning into the QD-dominated region with respect to the closest RRH with increasing UE transmit power. However, as the UE cluster moves away from the RRH, contributions from other RRHs gradually augment the achievable SINDR. Ultimately, when the UE cluster is equidistant from all the RRHs, each RRH contributes to the achievable SINDR in the same way, maximizing the achievable SINDR without dithering. Incorporating dithering significantly enhances UE performance, especially in proximity to any RRH, ensuring sustained maximum SINDR. However, adequate UE transmit power is essential to reach the farthest serving RRH. Consequently, the level of dithering required varies; more dithering is necessary when the UE cluster is close to any RRH, while no dithering is required when the UE cluster is centrally located. As expected, the max-min-SINDR with the BMMSE receiver exceeds that of the BMRC receiver in all the cases. However, in the case without dithering, the achievable max-min-SINDR is quickly improved with the BMMSE receiver compared to the BMRC receiver as the UE cluster moves to the center of the RRHs, since the BMMSE receiver better alleviates the QD compared to the BMRC receiver. 

                    \begin{figure}[t]
                    \centering
            \input{FigJor/fig10.tex}
             \vspace{-1mm}
            \caption{Max-min-SINDR versus area of the served UEs.} \label{fig:Disvsb}
            \vspace{-4mm}
        \end{figure}
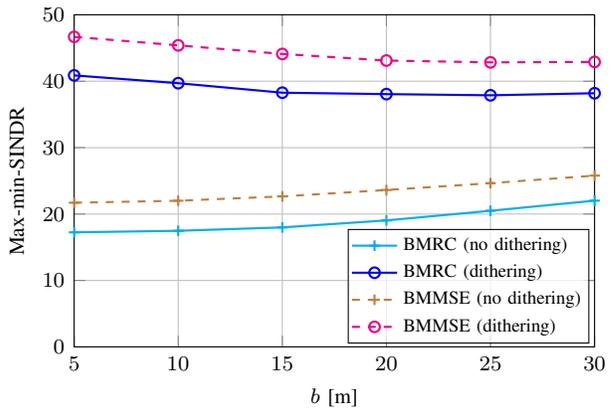

In Fig.~\ref{fig:Disvsb}, the max-min-SINDR is plotted against the UE cluster size $b$, with $d_{\textrm{ref}} =0$ and a maximum UE transmit power of $25$~dBm. As the UE cluster size increases, the effective received signal at the RRH is mostly dominated by a single UE. This results in: \textit{i)} a reduced SINDR gain from the joint reception across the RRHs with dithering, and \textit{ii)} an increase in the SINDR without dithering, due to reduced interference as shown in Fig.~\ref{fig:Disvsb}. However, even with a cluster size of $30$~m, dithering provides considerable performance gains with both the BMRC and BMMSE receivers.

  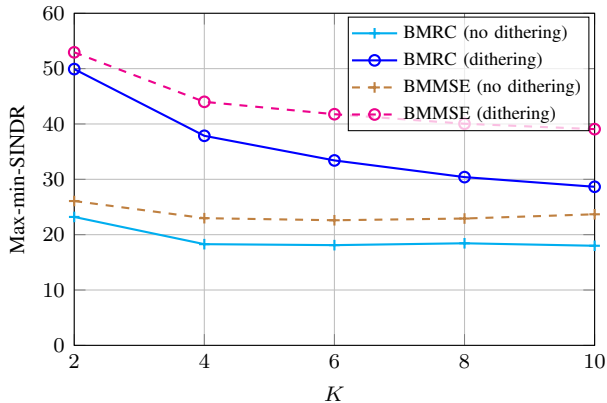
\begin{figure}[t]
  \centering
            \input{FigJor/fig11.tex}
             \vspace{-1mm}
            \caption{Max-min-SINDR versus number of UEs.} \label{fig:DisvsK}
            \vspace{-4mm}
        \end{figure}

In Fig.~\ref{fig:DisvsK}, the max-min-SINDR is plotted against the number of UEs for $d_{\textrm{ref}} = 0$ with a maximum UE transmit power of $25$~dBm. The UEs are placed on the perimeter of a circle with a radius of $10$~m, and all the UEs are equally spaced. With dithering, as the number of UEs increases, the max-min-SINDR decreases, mirroring the behavior observed in the unquantized systems. Furthermore, the BMMSE receiver demonstrates superior interference mitigation compared to the BMRC receiver, resulting in better performance gains as the number of UEs increases. In the case without dithering, the max-min-SINDR is not significantly impacted by increasing the number of UEs. This is because while the interference increases, it is observed that the correlated QD decreases with increasing the UEs.

          \begin{figure}[t]
          \centering
            \input{FigJor/fig12.tex}
             \vspace{-1mm}
            \caption{Max-min-SINDR versus number of RRH antennas.}
            \label{fig:DisvsM}
            \vspace{-4mm}
        \end{figure}
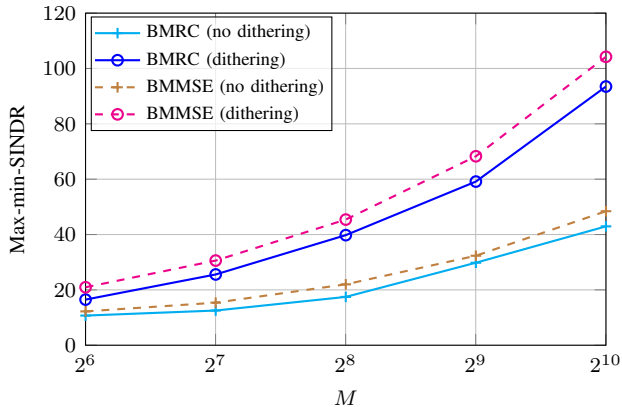
In Fig.~\ref{fig:DisvsM}, the max-min-SINDR is plotted as a function of the number of antennas at each RRH for $b=10$ and a UE transmit power of $25$~dBm. The max-min-SINDR increases with the increasing number of antennas at each RRH, both with and without dithering, similar to the behavior observed in the unquantized systems. It is observed that increasing the number of antennas at each RRH results in a decrease in the UE transmit power required to achieve the optimal max-min-SINDR. Additionally, with the increase in the number of antennas at each RRH, the gain from dithering is slightly better than in cases without dithering, since the QD increases with the number of antennas. Even with a high number of antennas at each RRH, there is still a considerable performance gap between the BMMSE receiver and the BMRC receiver due to the presence of the QD.

\begin{figure*}
\setcounter{equation}{42}
\begin{align}\label{eq:gradCrReal_bmrc}
\frac{\partial}{\partial \rho_k}\Re[\C_{\hat \r}(i,j)] & = \frac{\hat \A(i,i) \hat \A(j,j)}{\sqrt{1-|q_{i,j}|^2}} \bigg( \Re\big[\hat \h_k(i) \hat \h_k(j) + \C_{\tilde \h_k}(i,j)\big] - \frac{\pi}{4}\Re\big[\hat \C_{\y}(i,j)\big] \sum_{l \in \{i,j\}} \big( \big|\hat \h_k(l)\big|^2 + \C_{\tilde \h_k} (l,l) \big)\hat \A^{2}(l,l) \bigg)
\end{align}
\hrule
\setcounter{equation}{44}
\begin{align}\label{eq:p3gradCrReal_bmrc}
\frac{\partial}{\partial \sigma_b}\Re[\C_{\hat \r}(i,j)] = \frac{-\pi \sigma_b \hat \A(i,i) \hat \A(j,j)}{2\sqrt{1-|q_{i,j}|^2}} \Big( \Re\big[\hat \C_{ \y}(i,j)\big] \big( \hat \A^{2}(i,i) \E_b(i,i) + \hat \A^{2}(j,j) \E_b(j,j)\big)\Big)
\end{align}
\setcounter{equation}{41}
\hrule
\end{figure*}

\section{Conclusions}\label{sec:CONC}

We investigated the uplink SNDR of a UE in both single-RRH and multi-RRH scenarios with 1-bit ADCs. In a single-RRH case, the SNDR exhibits unimodal behavior with respect to the UE transmit power, whereas it becomes multimodal in a multi-RRH scenario. However, by adjusting the dithering at the RRHs, the SNDR can be optimized and made unimodal. In a multi-UE and multi-RRH scenario, considering the non-monotonic behavior of SINDR, we jointly optimized the UE transmit powers and RRH dithering levels for the min-power and max-min-SINDR objectives, employing both the BMRC and BMMSE receivers. For solving the UE transmit powers, we employed the gradient and BCD methods, while the line search method was utilized to determine the RRH dithering levels alongside the gradient method. Numerical results demonstrate that updating the UE transmit powers considering the RRH dithering enhances the system performance by enabling joint reception from multiple RRHs across several scenarios. Moreover, if achieving the desired target SNDR involves the reception from multiple RRHs using a large number of antennas and dithering, in such cases, the UE transmit power heavily depends on the distance to the farthest serving RRH.

\appendices

\section{Partial Derivative of $\hat \A$ and $\C_{\hat \r}$ with Respect to $\rho_k$}\label{app:rho_upd}

The gradient of $\hat \A$ with respect to $\rho_k$ is given by
\begin{align}
\frac{\partial}{\partial \rho_k}\hat \A = -\sqrt{\frac{1}{2\pi}}\Diag(\hat \C_{ \y})^{-\frac{3}{2}}\Diag(\hat \h_k \hat \h_k^{\herm} + \C_{\tilde \h_k}).
\end{align}
The gradient of the $(i,j)$th element ($i \ne j$) of the real part of $\C_{\hat \r}$ is given in \eqref{eq:gradCrReal_bmrc} at the top of the next page, with $q_{i,j} \triangleq \frac{\pi}{2} \C_{\hat \r}(i,j)$. \setcounter{equation}{43} The gradient of the imaginary part of $\C_{\hat \r}$ can be obtained by replacing $\Re[\cdot]$ with $\Im[\cdot]$ in \eqref{eq:gradCrReal_bmrc}, and $\frac{\partial}{\partial \rho_k}\C_{\hat \r}(i,j) = 0,~\forall i=j$.

\section{Partial Derivative of $\hat \A$ and $\C_{\hat \r}$ with Respect to $\sigma_b$}\label{app:grd_noise_min}

The gradient of $\hat \A$ with respect to $\sigma_b$ is given by
\begin{align}
\frac{\partial}{\partial \sigma_b}\hat \A = -2 \sigma_b\sqrt{\frac{1}{2\pi}}\Diag(\hat \C_{\y})^{-\frac{3}{2}}\E_b,
\end{align}
where $\E_b = \Diag \big( [c_1, \ldots, c_B] \big) \otimes \I_{M} \in \Compl^{BM \times BM}$ with $c_{\bar{b}} = 1$ if $\bar{b} = b$, otherwise $c_{\bar{b}} = 0$. The gradient of the $(i,j)$th element ($i \ne j$) of the real part of $\C_{\hat \r}$ is given in \eqref{eq:p3gradCrReal_bmrc} at the top of the next page. The gradient of the imaginary part of $\C_{\hat \r}$ can be obtained by replacing $\Re[\cdot]$ with $\Im[\cdot]$ in \eqref{eq:p3gradCrReal_bmrc}. Finally, we have $\frac{\partial}{\partial \sigma_b}\C_{\hat \r}(i,j) = 0,~\forall i=j$.

\bibliographystyle{IEEEtran}
\bibliography{refs_abbr,refs}
\end{document}

%% file: Definitions.tex
\renewcommand{\d}{\mathbf{d}}

\newcommand{\h}{\mathbf{h}}

\newcommand{\q}{\mathbf{q}}
\renewcommand{\r}{\mathbf{r}}

\newcommand{\w}{\mathbf{w}}

\newcommand{\y}{\mathbf{y}}
\newcommand{\z}{\mathbf{z}}

\newcommand{\0}{\mathbf{0}}

\newcommand{\A}{\mathbf{A}}

\newcommand{\C}{\mathbf{C}}

\newcommand{\E}{\mathbf{E}}

\newcommand{\I}{\mathbf{I}}

\newcommand{\setB}{\mathcal{B}}
\newcommand{\setC}{\mathcal{C}}

\newcommand{\setK}{\mathcal{K}}
\newcommand{\setL}{\mathcal{L}}

\newcommand{\setN}{\mathcal{N}}

\newcommand{\Compl}{\mbox{$\mathbb{C}$}}

\newcommand{\argmax}{\operatornamewithlimits{argmax}}
\newcommand{\argmin}{\operatornamewithlimits{argmin}}

\newcommand{\Diag}{\mathrm{Diag}}

\newcommand{\Exp}{\mathbb{E}}
\newcommand{\herm}{\mathrm{H}}
\renewcommand{\Im}{\mathrm{Im}}
\newcommand{\maximize}{\mathrm{maximize}}

\newcommand{\minimize}{\mathrm{minimize}}

\renewcommand{\Re}{\mathrm{Re}}
\newcommand{\sgn}{\mathrm{sgn}}
\newcommand{\st}{\mathrm{s.t.}}

\newcommand{\tran}{\mathrm{T}}

\newtheorem{theorem}{Theorem}
\newtheorem{definition}{Definition}
\newtheorem{lemma}{Lemma}

\newtheorem{remark}{Remark}


%% file: FigJor/fig2.tex
\begin{tikzpicture}

\begin{axis}[
	width=8cm,
	height=6cm,
	xmin=0, xmax=100,
	ymin=0, ymax=30,
    xlabel={Distance to the reference BS [m]},
    ylabel={$\mathrm{SNDR}$},
    ytick distance=5,
    xlabel near ticks,
	ylabel near ticks,
    x label style={font=\footnotesize},
	y label style={font=\footnotesize},
    ticklabel style={font=\footnotesize},
    legend style={at={(0.99,0.01)}, anchor=south east},
    legend cell align=left,
    legend columns=1,
    legend style={font=\scriptsize, inner sep=1pt, fill opacity=0.75, draw opacity=1, text opacity=1},
	grid=both,
]

\addplot[line width=1pt, black]
table[x=dist, y=SNDR_5, col sep=comma] 
{FigJor/data/fig2.txt};
\addlegendentry{$\rho_{k} = -5$~dBm};

\addplot[line width=1pt, red, dashed]
table[x=dist, y=SNDR_7p5, col sep=comma] 
{FigJor/data/fig2.txt};
\addlegendentry{$\rho_{k} = 7.5$~dBm};

\addplot[line width=1pt, blue, dotted]
table[x=dist, y=SNDR_20, col sep=comma] 
{FigJor/data/fig2.txt};
\addlegendentry{$\rho_{k} = 20$~dBm};

\end{axis}

\end{tikzpicture}

%% file: FigJor/fig3.tex
\begin{tikzpicture}

\begin{axis}[
	width=8cm,
	height=6cm,
	xmin=-20, xmax=60,
	ymin=0, ymax=30,
    xlabel={$\rho_{k}$ [dBm]},
    ylabel={$\mathrm{SNDR}$},
    ytick distance=5,
    xlabel near ticks,
	ylabel near ticks,
    x label style={font=\footnotesize},
	y label style={font=\footnotesize},
    ticklabel style={font=\footnotesize},
    legend style={at={(0.99,0.01)}, anchor=south east},
    legend cell align=left,
    legend columns=1,
    legend style={font=\scriptsize, inner sep=1pt, fill opacity=0.75, draw opacity=1, text opacity=1},
	grid=both,
]

\addplot[line width=1pt, black]
table[x=pwrdBm, y=UEat50, col sep=comma] 
{FigJor/data/fig3.txt};
\addlegendentry{UE at $50$~m (no dithering)};

\addplot[line width=1pt, red, dashed]
table[x=pwrdBm, y=UEat10, col sep=comma] 
{FigJor/data/fig3.txt};
\addlegendentry{UE at $10$~m (no dithering)};

\addplot[line width=1pt, blue, dotted]
table[x=pwrdBm, y=UEat10wTun, col sep=comma] 
{FigJor/data/fig3.txt};
\addlegendentry{UE at $10$~m (dithering)};

\end{axis}

\end{tikzpicture}

%% file: FigJor/fig5.tex
\begin{tikzpicture}

\begin{axis}[
	width=8.5cm,
	height=6cm,
	xmin=5, xmax=35,
	ymin=-15, ymax=25,
    xlabel={Target $\mathrm{SINDR}$},
    ylabel={Min-power [dBm]},
	ytick={-15,-5,5,15,25},
    xlabel near ticks,
	ylabel near ticks,
    x label style={font=\footnotesize},
	y label style={font=\footnotesize},
    ticklabel style={font=\footnotesize},
    legend style={at={(0.01,0.99)}, anchor=north west},
    legend cell align=left,
    legend columns=1,
    legend style={font=\scriptsize, inner sep=1pt, fill opacity=0.75, draw opacity=1, text opacity=1},
	grid=both,
]

\addplot[thick, purple, solid, mark=star, mark options={solid}]
table[x=SINDRAll, y=BMRCpower_gd, col sep=comma] 
{FigJor/data/fig5.txt};
\addlegendentry{BMRC (gradient)};

\addplot[thick, blue, solid, mark=square, mark options={solid}]
table[x=SINDRAll, y=BMRCpower, col sep=comma] 
{FigJor/data/fig5.txt};
\addlegendentry{BMRC (BCD)};

\addplot[thick, black, dashed, mark = star, mark options={solid}]
table[x=SINDRAll, y=BMMSEpower_gd, col sep=comma] 
{FigJor/data/fig5.txt};
\addlegendentry{BMMSE (gradient)};

\addplot[thick, magenta, dashed, mark=square, mark options={solid}]
table[x=SINDRAll, y=BMMSEpower, col sep=comma] 
{FigJor/data/fig5.txt};
\addlegendentry{BMMSE (BCD)};



\end{axis}

\end{tikzpicture}

%% file: FigJor/fig6.tex
\begin{tikzpicture}

\begin{axis}[
	width=8.5cm,
	height=6cm,
	xmin=-5, xmax=25,
	ymin=0, ymax=50,
    xlabel={$\rho_{\ue}$~[dBm]},
    ylabel={Max-min-SINDR},
    ytick distance=10,
    xlabel near ticks,
	ylabel near ticks,
    x label style={font=\footnotesize},
	y label style={font=\footnotesize},
    ticklabel style={font=\footnotesize},
    legend columns=2,
    legend style={at={(0.99,0.01)}, anchor=south east},
    legend style={font=\scriptsize, inner sep=1pt, fill opacity=0.75, draw opacity=1, text opacity=1},
	legend cell align=left,
	grid=both,
]

\addplot[thick, cyan,mark=+, solid , mark options={solid} ]
table[x=pwrdBm, y=BMRC_noDith, col sep=comma] 
{FigJor/data/fig6.txt};
\addlegendentry{BMRC (no dithering)};

\addplot[thick, blue, mark=o, solid, mark options={solid}]
table[x=pwrdBm, y=BMRC_dith, col sep=comma] 
{FigJor/data/fig6.txt};
\addlegendentry{BMRC (dithering)};

\addplot[thick, brown, mark=+, dashed , mark options={solid} ]
table[x=pwrdBm, y=BMMSE_noDith, col sep=comma] 
{FigJor/data/fig6.txt};
\addlegendentry{BMMSE (no dithering)};

\addplot[thick, magenta, dashed, mark=o, mark options={solid}]
table[x=pwrdBm, y=BMMSE_dith, col sep=comma] 
{FigJor/data/fig6.txt};
\addlegendentry{BMMSE (dithering)};

\addplot[thick, teal, dotted, mark=+, mark options={solid}]
table[x=pwrdBm, y=BMMSE_BMRCnodith, col sep=comma] 
{FigJor/data/fig6.txt};
\addlegendentry{Heuristic (no dithering)};

\addplot[thick, red, dotted, mark=o, mark options={solid}]
table[x=pwrdBm, y=BMMSE_BMRCdith, col sep=comma] 
{FigJor/data/fig6.txt};
\addlegendentry{Heuristic (dithering)};

\end{axis}

\end{tikzpicture}

%% file: FigJor/fig7.tex
\begin{tikzpicture}

\begin{axis}[
	width=8.5cm,
	height=6cm,
	xmin=-5, xmax=25,
	ymin=1e-4, ymax=0.1,
    xlabel={$\rho_{\ue}$~[dBm]},
    ylabel={SER},
    xlabel near ticks,
	ylabel near ticks,
    x label style={font=\footnotesize},
	y label style={font=\footnotesize},
    ticklabel style={font=\footnotesize},
     legend style={at={(0.01,0.01)}, anchor=south west},
    legend style={font=\scriptsize, inner sep=1pt, fill opacity=0.75, draw opacity=1, text opacity=1},
	legend cell align=left,
	grid=both,
 ymode=log,
    log basis y={10}
]

\addplot[thick, cyan,mark=+, solid , mark options={solid}  ]
table[x=pwrdBm, y=BMRCSERnoNoise, col sep=comma] 
{FigJor/data/fig7.txt};
\addlegendentry{BMRC (no dithering)};

\addplot[thick, blue, mark=o, solid, mark options={solid}]
table[x=pwrdBm, y=BMRCSERNoise, col sep=comma] 
{FigJor/data/fig7.txt};
\addlegendentry{BMRC (dithering)};

\addplot[thick, brown, mark=+, dashed , mark options={solid} ]
table[x=pwrdBm, y=BMMSESERnoNoise, col sep=comma] 
{FigJor/data/fig7.txt};
\addlegendentry{BMMSE (no dithering)};

\addplot[thick, magenta, dashed, mark=o, mark options={solid}]
table[x=pwrdBm, y=BMMSESERNoise, col sep=comma] 
{FigJor/data/fig7.txt};
\addlegendentry{BMMSE (dithering)};

\addplot[thick, teal, dotted, mark=+, mark options={solid}]
table[x=pwrdBm, y=BMRC_BMMSE_noNoise, col sep=comma] 
{FigJor/data/fig7.txt};
\addlegendentry{Heuristic (no dithering)};

\addplot[thick, red, dotted, mark=o, mark options={solid}]
table[x=pwrdBm, y=BMRC_BMMSE_Noise, col sep=comma] 
{FigJor/data/fig7.txt};
\addlegendentry{Heuristic (dithering)};

\end{axis}

\end{tikzpicture}

%% file: FigJor/fig8.tex
\begin{tikzpicture}

\begin{axis}[
	width=8.5cm,
	height=6cm,
	xmin=0, xmax=142,
	ymin=-10, ymax=25,
    xlabel={$d_{\textrm{ref}} + \frac{b}{\sqrt{2}}$ [m]},
    ylabel={Min-power [dBm]},
    ytick distance=5,
    xlabel near ticks,
	ylabel near ticks,
    x label style={font=\footnotesize},
	y label style={font=\footnotesize},
    ticklabel style={font=\footnotesize},
    legend cell align=left,
    legend columns=1,
    legend style={at={(0.5,0.01)}, anchor=south},
    legend style={font=\scriptsize, inner sep=1pt, fill opacity=0.75, draw opacity=1, text opacity=1},
	grid=both,
]

\addplot[thick, cyan,mark=x, solid , mark options={solid}  ]
table[x=disVal, y=PwrBMRC_tar15, col sep=comma] 
{FigJor/data/Fig8.txt};
\addlegendentry{BMRC (target $\mathrm{SINDR} = 15$)};

\addplot[thick, blue, mark=triangle, solid, mark options={solid}]
table[x=disVal, y=PwrBMRC_tar35, col sep=comma] 
{FigJor/data/Fig8.txt};
\addlegendentry{BMRC (target $\mathrm{SINDR} = 35$)};

\addplot[thick, brown, mark=x, dashed , mark options={solid} ]
table[x=disVal, y=PwrBMMSE_tar15, col sep=comma] 
{FigJor/data/Fig8.txt};
\addlegendentry{BMMSE (target $\mathrm{SINDR} = 15$)};

\addplot[thick, magenta, dashed, mark=triangle, mark options={solid}]
table[x=disVal, y=PwrBMMSE_tar35, col sep=comma] 
{FigJor/data/Fig8.txt};
\addlegendentry{BMMSE (target $\mathrm{SINDR} = 35$)};

\end{axis}

\end{tikzpicture}

%% file: FigJor/fig9.tex
\begin{tikzpicture}

\begin{axis}[
	width=8.5cm,
	height=6cm,
	xmin=0, xmax=142,
	ymin=15, ymax=50,
    xlabel={$d_{\textrm{ref}} + \frac{b}{\sqrt{2}}$~[m]},
    ylabel={Max-min-SINDR},
    ytick distance=5,
    xlabel near ticks,
	ylabel near ticks,
    x label style={font=\footnotesize},
	y label style={font=\footnotesize},
    ticklabel style={font=\footnotesize},
    legend style={at={(0.5,0.01)}, anchor=south},
    legend style={font=\scriptsize, inner sep=1pt, fill opacity=0.75, draw opacity=1, text opacity=1},
	legend cell align=left,
	grid=both,
]

\addplot[thick, cyan,mark=+, solid , mark options={solid} ]
table[x=disVal, y=BMRC_noNoise, col sep=comma] 
{FigJor/data/fig9.txt};
\addlegendentry{BMRC (no dithering)};

\addplot[thick, blue, mark=o, solid, mark options={solid}]
table[x=disVal, y=BMRC_Noise, col sep=comma] 
{FigJor/data/fig9.txt};
\addlegendentry{BMRC (dithering)};

\addplot[thick, brown, mark=+, dashed , mark options={solid} ]
table[x=disVal, y=BMMSE_noNoise, col sep=comma] 
{FigJor/data/fig9.txt};
\addlegendentry{BMMSE (no dithering)};

\addplot[thick, magenta, dashed, mark=o, mark options={solid}]
table[x=disVal, y=BMMSE_Noise, col sep=comma] 
{FigJor/data/fig9.txt};
\addlegendentry{BMMSE (dithering)};

\end{axis}

\end{tikzpicture}

%% file: FigJor/fig10.tex
\begin{tikzpicture}

\begin{axis}[
	width=8.5cm,
	height=6cm,
	xmin=5, xmax=30,
	ymin=0, ymax=50,
    xlabel={$b$~[m]},
    ylabel={Max-min-SINDR},
    ytick distance=10,
    xlabel near ticks,
	ylabel near ticks,
    x label style={font=\footnotesize},
	y label style={font=\footnotesize},
    ticklabel style={font=\footnotesize},
    legend style={at={(0.99,0.01)}, anchor=south east},
    legend style={font=\scriptsize, inner sep=1pt, fill opacity=0.75, draw opacity=1, text opacity=1},
	legend cell align=left,
	grid=both,
]

\addplot[thick, cyan,mark=+, solid , mark options={solid} ]
table[x=width, y=BMRC_noNoise, col sep=comma] 
{FigJor/data/fig10.txt};
\addlegendentry{BMRC (no dithering)};

\addplot[thick, blue, mark=o, solid, mark options={solid}]
table[x=width, y=BMRC_Noise, col sep=comma] 
{FigJor/data/fig10.txt};
\addlegendentry{BMRC (dithering)};

\addplot[thick, brown, mark=+, dashed , mark options={solid} ]
table[x=width, y=BMMSE_noNoise, col sep=comma] 
{FigJor/data/fig10.txt};
\addlegendentry{BMMSE (no dithering)};

\addplot[thick, magenta, dashed, mark=o, mark options={solid}]
table[x=width, y=BMMSE_Noise, col sep=comma] 
{FigJor/data/fig10.txt};
\addlegendentry{BMMSE (dithering)};

\end{axis}

\end{tikzpicture}

%% file: FigJor/fig11.tex
\begin{tikzpicture}

\begin{axis}[
	width=8.5cm,
	height=6cm,
	xmin=2, xmax=10,
	ymin=0, ymax=60,
    xlabel={$K$},
    ylabel={Max-min-SINDR},
    ytick distance=10,
    xlabel near ticks,
	ylabel near ticks,
    x label style={font=\footnotesize},
	y label style={font=\footnotesize},
    ticklabel style={font=\footnotesize},
  legend style={at={(0.99,0.99)}, anchor=north east},
    legend style={font=\scriptsize, inner sep=1pt, fill opacity=0.75, draw opacity=1, text opacity=1},
	legend cell align=left,
	grid=both,
]

\addplot[thick, cyan,mark=+, solid , mark options={solid} ]
table[x=UEall, y=BMRC_noNoise, col sep=comma] 
{FigJor/data/fig11.txt};
\addlegendentry{BMRC (no dithering)};

\addplot[thick, blue, mark=o, solid, mark options={solid}]
table[x=UEall, y=BMRC_Noise, col sep=comma] 
{FigJor/data/fig11.txt};
\addlegendentry{BMRC (dithering)};

\addplot[thick, brown, mark=+, dashed , mark options={solid} ]
table[x=UEall, y=BMMSE_noNoise, col sep=comma] 
{FigJor/data/fig11.txt};
\addlegendentry{BMMSE (no dithering)};

\addplot[thick, magenta, dashed, mark=o, mark options={solid}]
table[x=UEall, y=BMMSE_Noise, col sep=comma] 
{FigJor/data/fig11.txt};
\addlegendentry{BMMSE (dithering)};

\end{axis}

\end{tikzpicture}

%% file: FigJor/fig12.tex
\begin{tikzpicture}

\begin{axis}[
	width=8.5cm,
	height=6cm,
	xmin=64, xmax=1024,
	ymin=0, ymax=120,
    xlabel={$M$},
    ylabel={Max-min-SINDR},
    ytick distance=20,
    xlabel near ticks,
	ylabel near ticks,
    x label style={font=\footnotesize},
	y label style={font=\footnotesize},
    ticklabel style={font=\footnotesize},
    legend style={at={(0.01,0.99)}, anchor=north west},
    legend style={font=\scriptsize, inner sep=1pt, fill opacity=0.75, draw opacity=1, text opacity=1},
	legend cell align=left,
	grid=both,
 xmode=log,
    log basis x={2}
]

\addplot[thick, cyan,mark=+, solid , mark options={solid} ]
table[x=BSantAll, y=BMRC_noNoise, col sep=comma] 
{FigJor/data/fig12.txt};
\addlegendentry{BMRC (no dithering)};

\addplot[thick, blue, mark=o, solid, mark options={solid}]
table[x=BSantAll, y=BMRC_Noise, col sep=comma] 
{FigJor/data/fig12.txt};
\addlegendentry{BMRC (dithering)};

\addplot[thick, brown, mark=+, dashed , mark options={solid} ]
table[x=BSantAll, y=BMMSE_noNoise, col sep=comma] 
{FigJor/data/fig12.txt};
\addlegendentry{BMMSE (no dithering)};

\addplot[thick, magenta, dashed, mark=o, mark options={solid}]
table[x=BSantAll, y=BMMSE_Noise, col sep=comma] 
{FigJor/data/fig12.txt};
\addlegendentry{BMMSE (dithering)};

\end{axis}

\end{tikzpicture}